\documentclass{article}

\usepackage[sort&compress]{natbib}
\usepackage{arxiv} 
\usepackage{xtab,afterpage} 
\usepackage[printwatermark]{xwatermark} 
\usepackage{xcolor,pgf} 
\usepackage{longtable} 
\usepackage{multirow} 
\graphicspath{
	{images/}
}  
\DeclareGraphicsExtensions{.pdf,.jpeg,.png}
\usepackage{amssymb}
\usepackage{enumitem} 
\usepackage[linesnumbered]{algorithm2e} 
\usepackage[export]{adjustbox}

\setcitestyle{numbers,square}

\begin{document}
	
%
%

\title{Deep Learning in Information Security}

\author{
  Stefan Thaler, Vlado Menkovski\\
  Eindhoven University of Technology\\
  \AND
  Milan Petkovic \\
  Eindhoven University of Technology, Philips Research Laboratories\\
}

\maketitle

%
%
%
%
\begin{abstract}
Machine learning has a long tradition of helping to solve complex information security problems that are difficult to solve manually. Machine learning techniques learn models from data representations to solve a task. These data representations are hand-crafted by domain experts. Deep Learning is a sub-field of machine learning, which uses models that are composed of multiple layers. Consequently, representations that are used to solve a task are learned from the data instead of being manually designed.

In this survey, we study the use of DL techniques within the domain of information security. We systematically reviewed 77 papers and presented them from a data-centric perspective. This data-centric perspective reflects one of the most crucial advantages of DL techniques -- domain independence. If DL-methods succeed to solve problems on a data type in one domain, they most likely will also succeed on similar data from another domain. Other advantages of DL methods are unrivaled scalability and efficiency, both regarding the number of examples that can be analyzed as well as with respect of dimensionality of the input data. DL methods generally are capable of achieving high-performance and generalize well.

However, information security is a domain with unique requirements and challenges. Based on an analysis of our reviewed papers, we point out shortcomings of DL-methods to those requirements and discuss further research opportunities.
\end{abstract}

%
%
%
%
\section{Introduction}
\label{sec:introduction}

Information security (InfoSec) addresses the protection of information and information systems from unauthorized access, use, disclosure, disruption, modification, or destruction in order to provide confidentiality, integrity, and availability~\citep{barker2003guidelines}. Many InfoSec challenges involve the analysisof large amounts of data. For some of the challenges, it is impractical to analyze the data manually or to write software, because either the volume of the data is too large, or formalizing all the knowledge a computer needs to know to solve the problem is challenging task itself. To overcome such problems, many methods that are based on machine learning have been proposed~\citep{joseph_et_al:DM:2013:4356}. Machine learning techniques are computer algorithms that learn a model from experience to solve a task according to a performance measure~\citep{mitchell1997machine}. Examples for tasks that can be addressed with machine-learning are classification, regression or anomaly detection. 

Deep Learning (DL) is a sub-field of machine learning. DL techniques use models that consist of multiple layers of abstraction. In contrast to traditional machine learning, DL models can learn representations that are useful for solving a task from data, instead of relying on expert-designed representations. Commonly, the 'deep' refers to the depth of the model regarding layers. In the past few years, an increase of available data and computing power led to remarkable successes of DL techniques in many domains, e.g., image classification~\citep{krizhevsky2012imagenet}, speech recognition~\citep{hinton2012deep} or end-to-end machine translation~\citep{bahdanau2014neural}.%

DL algorithms have many properties that make them attractive for solving InfoSec problems. %
First, DL techniques scale very well to large amounts of training data and also scale well with respect to the number of model parameters~\citep{shazeer2017outrageously}. Scalability is of great interest in InfoSec, since the amount of data is produced and needs to be analyzed grows exponentially. The amount of data that is worldwide produced is projected to grow to to 163 zettabytes in 2025~\citep{Reinsel2017}. 
Secondly, DL techniques are in general capable of learning from high-dimensional data. In contrast, traditional ML algorithms suffer from the curse of dimensionality~\citep{keogh2011curse}. That is, high-dimensional input data is problematic for methods that require statistical significance to learn useful properties from the data. 
Furthermore, DL models are capable of learning distributed representations, which can be exponentially more expressive than non-distributed representations. Distributed representations are composed of many elements that can be set for each other. This is powerful because they can use $n$ features with $k$ values to represent $n^k$ concepts~\citep{bengio2013representation}. 
Finally, since DL models can learn representations from data, they reduce the need for feature engineering. Many shallow machine learning approaches need hand-crafted features to solve a task. Such feature engineering is labor-intensive and hand-crafted features are often brittle and problem specific. The capability to learn features from data also allows to transfer solutions from one domain to another, as long as the underlying data-type is the same.  

DL algorithms have many compelling properties and have recently become hugely successful in other domains. Hence, in this literature review we study the application of DL technologies on InfoSec problems. We aim to understand how DL can be used tackle InfoSec problems, which tasks have been addressed and which challenges are remaining. In detail, our contributions are:
\begin{itemize}
  \item We conduct a systematical literature review on DL research in InfoSec. We conduct this review from a data-centric perspective. We propose to classify the papers along three dimensions: data-type, task, and model. We present our classification scheme in Section \ref{sec:classification-scheme}.  We review 66 papers that apply DL on five different data-types: sequential data in Section \ref{sec:sequential-data}, spatial data in Section \ref{sec:spatial-data}, structured data in Section \ref{sec:structured-data}, text data in Section \ref{sec:text-data} and multi-modal data in Section \ref{sec:multi-modal-data}. Additionally, we present 11 papers, that study security properties of DL algorithms such as privacy and integrity in Section \ref{sec:sec-properties-dl-algorithms}.
  \item We discuss the application in current approaches in Section \ref{sec:discussion} and highlight their strength. We then touch on special requirements in the InfoSec domain and outline future challenges and research opportunities for DL in InfoSec.  
\end{itemize}

The remaining paper is structured as follows. In Section \ref{sec:background}, we introduce the necessary background for this paper. We first briefly state the fundamental goals of InfoSec and the role of machine learning in InfoSec~(Section \ref{sec:background-information-security}). Then, we give an overview over the fundamental components of DL techniques~(Section \ref{sec:background-deep-learning}), which is followed by an overview of the common DL architectures (Section \ref{sec:deep-learning-models}) , data types~(Section \ref{sec:bg:data-types}) and tasks~(Section \ref{sec:bg:tasks}). Thereafter, we introduce our classification scheme (Section~\ref{sec:classification-scheme}) and detail our survey methodology~(Section \ref{sec:survey-methodology}). We present the actual survey in Section~\ref{sec:survey}. Section~\ref{sec:survey} consists of five sub-sections, one for each data type that we distinguish. Reviewed papers are grouped first according to the data-type that is analyzed, and then according to the task that is addressed. In Section \ref{sec:discussion} we discuss challenges and outline future research directions for DL in InfoSec. 

%
%
%
%
\section{Background}
\label{sec:background}

\subsection{Information Security}
\label{sec:background-information-security}
InfoSec is the field of protection of information and information systems from unauthorized access, use, disclosure, disruption, modification, or destruction to provide confidentiality, integrity, and availability~\citep{kissel2013glossary}. 

InfoSec is structured around fundamental security objectives~\citep{stallings2012computer}
: {\it Data confidentiality}, which assures that private information is not made available or disclosed to unauthorized individuals; {\it Privacy}, which assures that individuals control or influence what information related to them may be collected and stored and by whom and to whom that information may be disclosed; {\it Data integrity}, which assures that information and programs are changed only in a specific and authorized manner; {\it System integrity}, which assures that a system performs its intended function in an unimpaired manner, free from deliberate or inadvertent unauthorized manipulation of the system; {\it Availability}, which assures that systems work promptly, and service is not denied to authorized users; {\it Authenticity}, which assures that users are who they claim to be and that each input arriving at the system came from a trusted sources; and {\it Accountability}, which assures that actions of an entity are traceable uniquely to that entity.

Most problems in InfoSec are complex and challenging to tackle for many reasons. The assets that need to be continuously protected evolve and are often organized into large, complex systems. The attackers continuously develop new methods for attacking systems. Hand-crafted or statistically derived mechanism to address modern-day security problems become increasingly difficult to design and are labor intensive to maintain. Machine-learning techniques allow computer programs to solve InfoSec challenges by learning solutions from data, which would otherwise be difficult to solve. Joseph et al. provide an overview over the use of Machine Learning for InfoSec~\citep{joseph_et_al:DM:2013:4356}. 

\subsection{Deep Learning}
\label{sec:background-deep-learning}

\begin{figure}[ht]
\begin{center}
\centerline{\includegraphics[width=0.70\columnwidth]{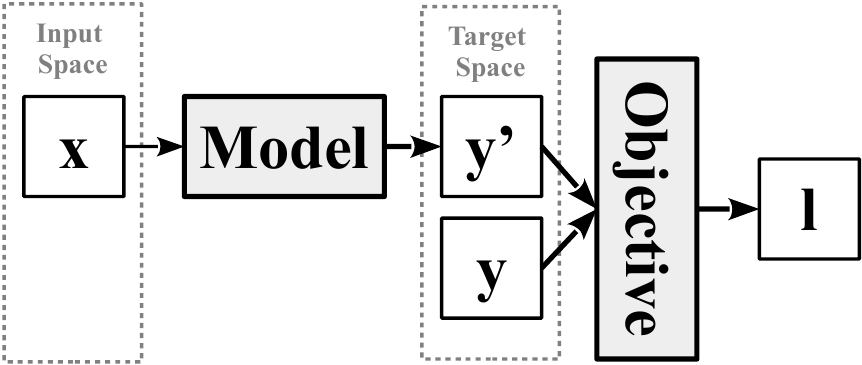}}
\caption{DL overview. A model, which is a parametrized function, maps data $x$ from input into target space $y'$. An objective function and a ground truth $y$ is used to calculate a loss $l$. The loss measures of how 'wrong' the model predicted the $y$ given $x$.}
\label{fig:bg:dl-overview}
\end{center}
\end{figure}

DL is a fast-growing field, and a comprehensive overview is out of scope for this paper. Instead, in this section we want to briefly introduce  the main components of a DL algorithms. DL-based methods commonly consist of four components: a model, data, an objective and an optimization procedure. We sketch the relationship between data, model and the objective in Figure \ref{fig:bg:dl-overview}. In the following paragraphs, we will introduce these four components. 

\paragraph{Model}
The model is a parameterized function, which maps from the input data, which often has a high dimensionality, to target data, which often has lower dimensionality. An optimization procedure learns the parameters of such a model. Furthermore, a model is often composed in different ways, commonly referred to as architecture. We present different ways on how to construct such composite 'models in Section \ref{sec:deep-learning-models}. 

\paragraph{Data}
Data is an essential ingredient of DL-approaches. Since DL methods learn useful representations to solve a task from data, the learned model can only be as representative as the data was of the problem that is about to be solved. If the data is not representative, the model most likely is also not. Mostly, data used to train DL-models is high dimensional. One of the fundamental assumptions is that such high dimensional data concentrates around a low dimensional manifold, and the DL algorithm is capable of finding such lower dimensional manifolds. 

\paragraph{Objective}
The objective is a function that defines how the parameters are learned. It tells a model the prediction error towards the ground truth. It is often also referred to as goal-, cost-, or loss-function. An example objective is the Mean Squared Error. 

\paragraph{Optimization procedure}
The optimization procedure describes the steps to perform the actual learning. In DL, the optimization procedure is almost always a form of mini-batch stochastic gradient descent. Mini-batch stochastic gradient descent performs the following steps for a subset of the overall data: a forward pass, which calculates the predictions $y'$ given a model and the data $x$; a backward pass, which calculates the gradients of all parameters with respect to to the objective; an parameter update step, which updates the parameters of the model. More recently, the DL community switched to momentum-based forms of mini-batch SGD such as Adam~\citep{Kingma2014} or RMSprop~\citep{tieleman2012lecture}. 

For a more comprehensive overview over DL, we would like to point to excellent resources that attempt to do so. J{\"u}rgen Schmidhuber provides a historical overview over the field up to now ~\citep{schmidhuber2015deep}. In this overview, he summarizes popular algorithms and chronologically attributes them. Bengio et al. provide an overview of the DL field~ \citep{bengio2009learning}, and later with a focus on the unsupervised, representation learning approaches~\citep{bengio2013representation}. Furthermore, Deng et al. present another overview over DL methods and applications~\citep{deng2014deep}. Goodfellow et al. published a textbook on DL~\citep{Goodfellow-et-al-2016-Book}, which is also available online~\footnote{http://www.deeplearningbook.org/}. 

%
%
\subsection{Deep Learning Architectures}
\label{sec:deep-learning-models}

DL models are composed of different "building blocks" -- often called layers -- and the composition of these building blocks is commonly referred to as model architecture. Mathematically speaking, a layer is a function $y=f(x)$ that maps some input $x$ to some output $y$, mostly $f$ is a parameterized function. Here we list and briefly describe common DL model building blocks. Describing them or even describing their functioning in a detailed way is out of the scope of this survey paper, especially since many of the technologies rapidly evolve and adapt. Instead of providing details, we attempt to explain these building blocks on a high level, highlight common use cases and reference to important developments of the field. In this section, we will first introduce common layers, and then introduce a few popular composite architectures.

As the term "building block" hints, DL building blocks can be arbitrarily combined. Three examples of such building blocks are convolutional layers and recurrent layers and fully connected layers. The model architect can combine these building blocks into to a model for a sequence analysis task, where the convolutional layer reduces the input dimension of the sequence and abstract local, spatial correlation, the recurrent layer models temporal connections and the fully connected layer models complex relationships of the temporal model that are useful for the analysis task. 

Which building block, how many of them, and how to combine them often depends on the data and the task at hand, and depends on the researcher addressing the task. Decisions often involve domain knowledge and knowledge about the data structure. A few guidelines and best practices exist~\citep{montavon2012neural}, such as random search for hyper-parameter optimization~\citep{bergstra2012random}, advice for restricted Boltzmann machines~ \citep{hinton2012practical}, tips for using stochastic gradient descent~\citep{bottou2012stochastic}, design of deep architectures~\citep{bengio2012practical}, or how to calculate the size of the receptive field of convolutional neural networks~\citep{dumoulin2016guide}.

\begin{figure}[ht]
\begin{center}
\centerline{\includegraphics[width=0.70\columnwidth]{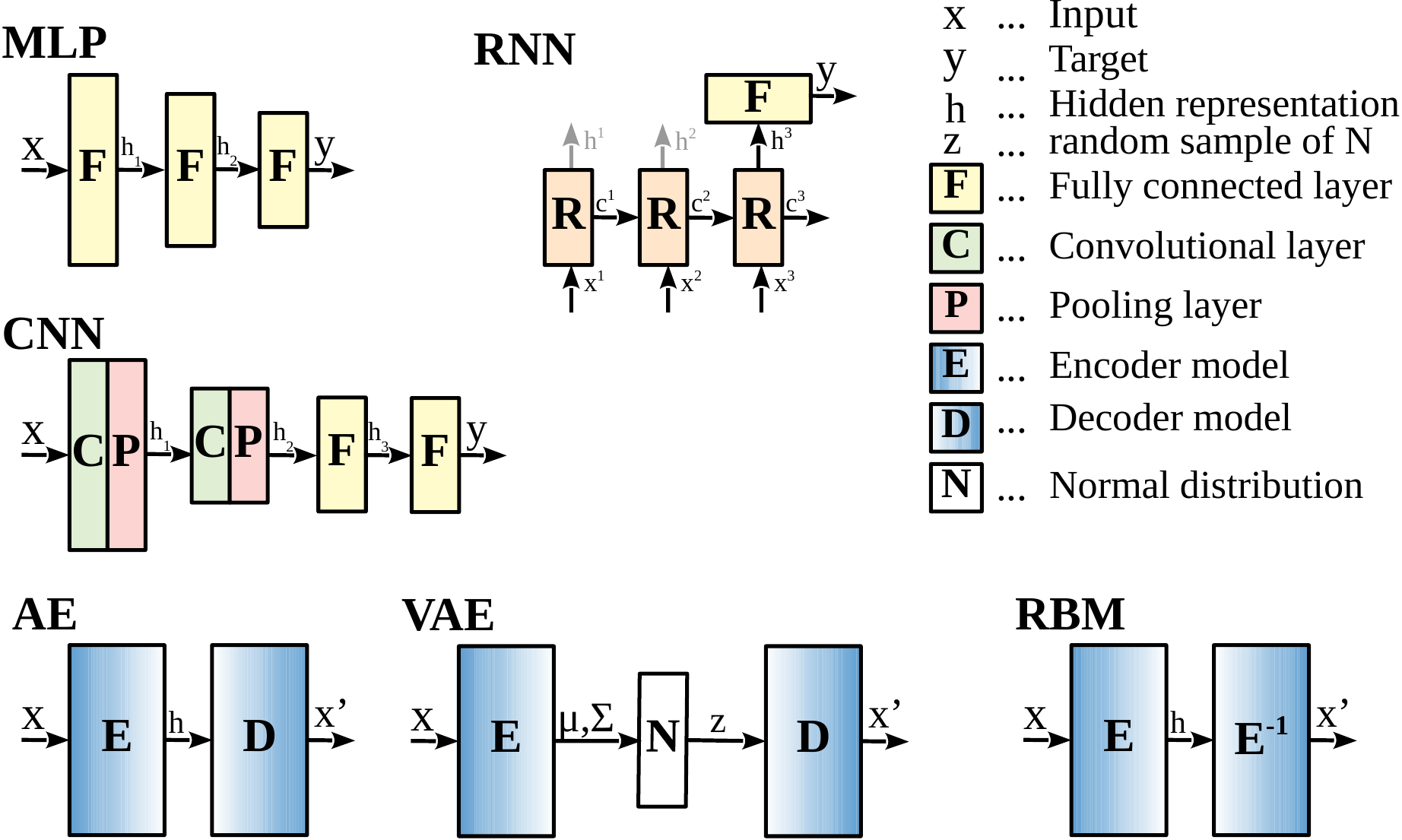}}
\caption{Schematics of common DL composite architectures. Architectures sketched are Multi-layer perceptron (MLP), Convolutional Neural Network (CNN), Recurrent neural network (RNN), Autoencder (AE), Variational Autoencoder (VAE) and a restricted Boltzmann machine (RBM).}
\label{fig:architecture-types-schematic}
\end{center}
\end{figure}

We provide an overview of common composite models in Figure \ref{fig:architecture-types-schematic}. There are many other model architectures, for example, Generative Adversarial Networks (GANs)~\citep{goodfellow2014generative}, or deep Boltzmann Machines (DBMs)~\citep{salakhutdinov2010efficient}, as well as countless variations, combinations, and improvements of different layers, e.g., attention mechanisms for RNNs~\citep{bahdanau2014neural}. A comprehensive overview is out of scope, but many models, techniques, and recent developments can be found in Bengio et al.'s survey\citep{bengio2013representation}  and Juergen Schmidhuber's~\citep{schmidhuber2015deep}. 

\paragraph{Fully-connected Layers}
A fully-connected (FC) layer connects each of the inputs elements to each of the output elements in a non-linear way, i.e., $y=\sigma(Wx+b)$, with $\sigma$ being a non-linear activation function. Fully-Connected layers are a fundamental building block and can be found in many DL model architectures. Multiple, chained fully-connected layers can learn arbitrary function mappings between the input an the output~\citep{hornik1991approximation}. 

\paragraph{Convolutional Layers}
Convolutional layers (CO) use locally shared parameters to learn an activation map of an input. The activation map is derived by "sliding" a kernel over the input and calculating a dot product of the kernel and the input segment. Often, an activation map is passed through a non-linear activation function. Convolutional layers are fundamental building blocks for convolutional neural networks and used to learn hierarchically structured features, especially in spatial data.

\paragraph{Pooling Layers}
Pooling layers (PO) reduce the size of representations, and thereby achieve desirable properties such as translation- or shift- invariance. Common ways to achieve that reduction are taking the maximum~\citep{riesenhuber1999hierarchical} or the sub-sampling layers~\citep{lecun1998gradient}. Pooling layers are often found in convolutional neural networks. 

\paragraph{Recurrent Layers}
Recurrent layers learn relationships about the input data by having a recursive function defined over different parts of the input. LSTMs~\citep{hochreiter1997long} and GRUs~\citep{cho2014learning} are among the most popular recurrent layers. Recurrent layers are building blocks for recurrent neural networks which are useful for many tasks in sequential and text data, where a state of the sequence plays an important role.

\paragraph{Restricted Boltzmann Machines}
Restricted Boltzmann Machines are energy-based models that learn a joint probability distribution of the input and output data~\citep{smolensky1986information}. An energy function defines this joint probability distribution. The restriction in the Restriction Boltzmann Machine is that there are no intra-layer connections in the hidden layer. 

\paragraph{Dropout Layers}
Dropout layers are parameterless layers, which stochastically add noise to the input or set the input to zero~\citep{krizhevsky2012imagenet, srivastava2014dropout}. Dropout essentially turns one network into an ensemble of networks, thereby increasing the generalization capability of the network. 

These layers can be composed to more sophisticated models. Here, we briefly introduce six common model architectures.

\paragraph{Multi-Layer Perceptrons (MLP)}
Multi-layer Perceptrons (MLPs) are feed-forward neural networks that are composed of multiple, fully-connected layers. This composition allows each layer to use the features of the previous layer to create more abstract features. Such a network learns to produce distributed representations that help to solve tasks efficiently. MLPs can approximate arbitrary, non-linear functions, and should not be confused with perceptrons, which can only learn simple, linear relationships between the input and the target data. One significant disadvantage of MLPs is that the model size grows proportionally with respect to the dimension of the input features. 

\paragraph{Convolutional Neural Networks (CNN)}
Feedforward neural networks make minimal assumptions about the data that they are processing. However, often we have general information about the data that we are processing. One example of such data is images. The pixel in images usually have a loose spatial correlation, i.e., a pixel will loosely influence the value of the pixels in its vicinity. 

Convolutional neural networks (CNNs) are a particular type of feed-forward neural network that use this spatial information correlation to design neural networks that are better for processing such data. CNNs combine three key ideas: local receptive fields, parameter sharing, and local sub-sampling. 

Local receptive fields, also called kernels, connect small patches of the input data with one point of the output data. Local receptive fields assume that the input data are spatially correlated, i.e., that the neighborhood of a data point influences this data point and vice versa. Connecting all small patches with own parameters to all outputs would be computationally impractical as it would increase the number of parameters to learn required drastically. Instead, parameters for such local receptive fields are "slid" over the input, and an output is calculated for each different position. The parameters are shared for each position. When training CNNS, multiple kernels per CO layer will be trained and slid over the input, thereby producing multiple activation maps. This approach drastically reduces the number of parameters needed. 

CNNs are typically composed of multiple convolution layers, pooling layers, and fully-connected layers. A combination of CO and PO layers learn hierarchical representations from the data, and the FC layers learn complex interdependencies between such representations. Examples of such architectures are AlexNet~\citep{krizhevsky2012imagenet} and VGG-Net~\citep{simonyan2014very}. Other compositions are possible, such as a CNN that consists purely of CO-layers~\citep{springenberg2014striving}.

\paragraph{Recurrent Neural Networks (RNN)}
%
Another general assumption one can make about the data is temporal (or sequential) interdependence of data, as time series, natural language or sound. If the data is locally sequentially correlated, 1-D convolutional neural networks can be used to learn representations from such patterns. For more complicated patterns or patterns that occur over time, recurrent neural networks (RNNs) have been developed. 

RNNs are neural networks that are designed in a way to reuse the outputs of the network in later calculations, which also clarifies the name. Similarly to CNNS, recurrent neural networks rely on parameter sharing, but in a different fashion. In addition to parameter sharing, recurrent neural networks remain a state (or context) of the network, which they pass on for further processing. 

RNNs can be used to learn functions in flexible ways. They can be used to learn functions to map sequences to a single output (many-to-one), for example, to classify sequences. They can be used to learn functions that map sequences to other sequences (many-to-many), for example, to tag sequences with specific labels or for natural language translation~\citep{cho2014learning}. Moreover, they can be used to map a single value to multiple outputs (one-to-many), for example, to generate descriptive text from an input image~\citep{vinyals2015show}. 

RNNs are typically composed of one or more recurrent layers (RE) to represent the sequences and one or more fully connected layers (FC) to learn complex relationships for such sequence representations. A common variant of the RNN is the bi-directional RNN, which combines two RE layers, one that processes the sequence of inputs from start to end and the other in reverse order. The output of these two layers is subsequently merged. 

\paragraph{Autoencoders (AE)}
Autoencoders (AE) are composite models that consist of two components: an encoder model and a decoder model. The task of an AE is to output a reconstruction of the input under certain constraints. The encoder and the decoder Model can be any Neural Network, preferably one that works well with the data type. So one can imagine an AE for images where the component models are CNNs or an AE for text where the component models are RNN. 

If AE have sufficient capacity, they will learn two functions that will copy the input to the output. Such functions are generally not useful. Therefore the representations that the AE has to learn are typically constrained in specific ways, for example by sparsity, or by a form of regularization. Such constraints force the encoder model to learn representations that contain potentially useful properties or regularities of the data. One use case for AEs is dimensionality reduction.

\paragraph{Variational Autoencoders}
Similar to an autoencoder, the Variational Autoencoder (AE) consist of two models, an encoder model, and a decoder model. The encoder model learns $\Sigma$ and $\mu$ of a multivariate Gaussian distribution given a particular input $x$. This distribution is used to sample a random variable $z$. The decoder model learns to reconstruct $x$ given $z$. The decoder model is the generative model.

\paragraph{Deep Belief Networks (DBN)}
Deep Belief Networks (DBNs) are generative models that are composed of multiple layers of restricted Boltzmann machines~\citep{hinton2006fast}. Each layer is trained individually and then combined to form a DBN. DBNs have mostly been used to learn input representations in an unsupervised way, for example, \citep{vincent2008extracting}. DBNs are trained by greedily training each component RBM in an unsupervised way. DBNs are generative models, so they can also be used to generate samples. DBNs and RBMs have mostly been replaced by other techniques~\citep{Goodfellow-et-al-2016-Book}. 

%
%
\subsection{Data Types}
\label{sec:bg:data-types}
DL algorithms learn models from data. Data can be organized in many forms. The organization depends on the domain and the task that is solved. There is no one right way to organize, but commonly data are organized in a way that is efficient to process or natural for the domain. In this survey, we distinguish between five different data types: spatial data (PD), sequential data (QD), and structured data (SD), text data (TD) and multi-modal data (MD). In the next paragraphs, we will describe these five data types in more detail.

\paragraph{Spatial Data}
Spatial data is data that is organized in such a way that individual data points are addressable by an N-dimensional coordinate system. For certain data types, the spatial location of data points carries much information about the information stored in the data. An example is images, where much of the information that is contained in the image is stored the location of a specific pixel. Consequently, a spatial representation of such data offers the advantage of leveraging the information that is stored in this information. 

In InfoSec, the most prevalent type of spatial data which DL is applied to is images. Images occur in naturally in domains such as biometrics, steganography, or steganalysis where images are a data modality of central interest for the domain. However, images also occur in other domains, such as in website security, where researchers attempt to break CAPTCHAs, which are often visual. Apart from that, only a few approaches in InfoSec use DL on spatial data, such as organizing audio data into a spatial arrangement using spectrograms, or binary malware data that gets transformed into a spatial arrangement. 

\paragraph{Sequential Data}
Generally, sequential data are ordered lists of events, where an event can be represented as a symbolic value, a real numerical value, a vector of real or symbolic values or a complex data type~\citep{xing2010brief}. Sequential data can also be considered as 1D spatial data, that is, locally correlated and each event is referable via a 1D coordinate system. Furthermore, text can also be seen as sequential data, either as a sequence of characters or as a sequence of words, although the relationship between the elements of text is complex. The distinguishing is arbitrary and depends on the data, the treatment of the data and the chosen data analysis techniques. 

In this survey, we treat sequential data as sequential, if it is analyzed with respect to the sequential nature of the data. For example, network traffic can be treated as a sequence of packets, but also each packet could be analyzed individually, without regard to the order of the packets. 
Many data types in InfoSec can be treated as sequential data. In our survey, we have three main categories of sequential data: operation codes, network traffic, and audio data. 

\paragraph{Structured Data}
Structured data are data whose creation depends on a fixed data model, for example, a record in a relational database. Semi-structured is a data type that has a structured format, but no fixed data model, for example, graph data or JSON formatted data. In this survey, we combine structured and semi-structured data, since from a DL perspective they are very similar.   

A prominent source for structured data in InfoSec is network traffic. Network traffic is usually parted into packages, each of which contains a fixed set of fields with specific values. Another source of structured data is event logs that are stored in a database. We also consider data where a fixed set of high-level features (the stress is on high-level) is extracted from data as structured data because this equals a record in a database where each field of the entry corresponds to a feature. 

Among the surveyed paper, we found three tasks on structured and semi-structured data: code similarity detection, network intrusion detection, and drive-by attack detection. 

\paragraph{Text Data}

Text data are a form of unstructured, sequential data. Commonly, a text is interpreted in two ways: as a sequence of characters, or as a sequence of tokens where tokens can be words, word-parts, punctuation or stopping characters. The elements of a time series are generally related to each other in a linear, temporal fashion, i.e., elements from earlier time steps. This relationship is often also true for text data. However, the relationships between the elements are more complicated, often long-term and high-level. 

In InfoSec, text data is mainly found in log files and in analyzing communication such as social media messages or emails. Another source of text may be the source code of software to identify vulnerabilities. 
 
Text data is generally treated as sequential data and analyzed with similar means. One problem of analyzing text is to find suitable representations of the input words. Text data is high dimensional, which renders sparse representations computationally impractical. Hence, commonly dense representations such as Word2Vec~\citep{NIPS2013_5021} and GloVe~\citep{pennington2014glove} are used to represent input words. 

\paragraph{Multi-Modal Data}
At times more than one data modality is used for solving a problem, or multiple modalities are derived from the same data type. An example of multiple modalities are videos that contain both a spatial and a temporal component. Such data can be analyzed by combining models like CNNs and LSTMs to represent both of the aspects of the data. 

In this survey we only have two examples of multi-modal data, one that treats ECG data as spatial as well as sequential data, and 
moreover, one that uses text and images for the detection of cyberbullying.

%
%
\subsection{Tasks}
\label{sec:bg:tasks}

Tasks are the problems that should be solved. One of the key challenges to solve a problem with machine learning is how to present the problem in such a way that the machine learning algorithm can solve it. For example, the problem of finding malware can be reduced to a classification problem, i.e., given a binary file, decide whether it is malicious or benign. How to solve a problem and which task to use is often a design decision, since the borders between different tasks are often vague or a problem can be solved in multiple ways. For instance, malware could be detected by a classifier that classifies binaries into malicious or benign. Alternatively, malware could also be detected by clustering, which groups malicious and benign binaries close together. 

Here we describe four common tasks, anomaly detection, classification, clustering, dimensionality reduction and representation learning. This selection is by no means a complete, as there are many other tasks such as regression, machine translation, imputation of values, manifold learning or metric learning. We do not describe all of them since up to now many of them do not play a significant in the papers that we have reviewed. 

\paragraph{Anomaly detection (AD)} 
Anomaly detection (AD) is the task of searching for atypical objects or events from a larger group. AD techniques have frequently been applied to InfoSec problems such as intrusion detection, malware detection or insider threat detection. DL-techniques recently were also explored for anomaly detection, e.g., for detecting anomalous log lines~\citep{Du2017}. There are additional considerations when using anomaly detection to solve an InfoSec problem. For example, when used for intrusion detection, one needs to show that anomalous data is actually malicious~\citep{gates2006challenging}. 

\paragraph{Classification}
Classification is the task of mapping an input sample to a discrete category of output classes. Many problems can be formulated as a classification task, and often in multiple ways. For example, biometric fingerprint matching can be framed as a classification task in two ways. One could frame it as binary classification task of given two fingerprints, are they from the same individual. Alternatively, one could ask: given a fingerprint, to which individually does it belong. Regression tasks are closely related to classification tasks, but instead of discrete output categories, the goal is to predict a numerical value. 

\paragraph{Clustering}
Clustering (CL) is the task of finding groups and group membership in a set of objects. Objects that are similar to a certain metric and attributes should be grouped, and objects that are dissimilar should not be grouped. Clustering is related to classification in the sense that it assigns membership to a certain group. However, the main focus of clustering is to find the groups. Deciding the number of groups is a currently unsolved problem, but several heuristics exist~\citep{sugar2003finding,tibshirani2001estimating}. 

In InfoSec, clustering techniques have the appealing property that they deal well with unknown groups and unknown objects. However, same problems exist as for anomaly detection. In InfoSec, DL-techniques are not directly used for clustering but instead used to learn representations that can in a second step be clustered. 

\paragraph{Representation learning}
Representation learning (RL) is the task of finding another representation for some input data. When using DL-techniques for representation learning, representations are always learned, because the architecture usually consists of multiple layers of representation. For example, when classifying a binary file into malicious or benign, the model that is being trained automatically learns a representation of the data. However, in some cases, no other explicit tasks is addressed, for example when an AE is used. There, the primary focus is to find a representation for the data that fulfills specific properties, e.g., a reduction of input dimension or a denoised version of the input. 

In InfoSec, explicit representation learning using DL-techniques is often used to reduce the input dimensions and in combination with another classifier~\citep{krizhevsky2012imagenet,Luo2017} or to use domain knowledge from a similar task, e.g., ~\citep{Wang2015}.

%
%
%
%
\section{Classification Scheme}
\label{sec:classification-scheme}

This survey aims to understand how DL techniques have been applied to InfoSec problems. DL methods are data-centric but mainly domain-independent methods. Hence we present this survey from a data-centric perspective, not from a topic-centric perspective. We chose this perspective because it allows researchers from other InfoSec sub-domains and other non-security domains that have a problem with similar data types see how these problems were addressed, and potentially to reuse solutions from other fields.

\begin{figure}[htbp]
\begin{center}
\centerline{\includegraphics[width=0.70\columnwidth]{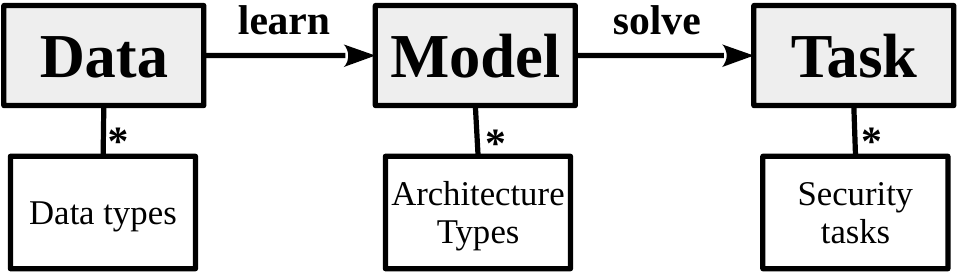}}
\caption{We use data centric perspective on DL in InfoSec. We learn a model using data to solve a problem.}
\label{fig:cat-overview}
\end{center}
\end{figure}

\begin{figure}[htbp]
\begin{center}
\centerline{\includegraphics[width=0.9\columnwidth]{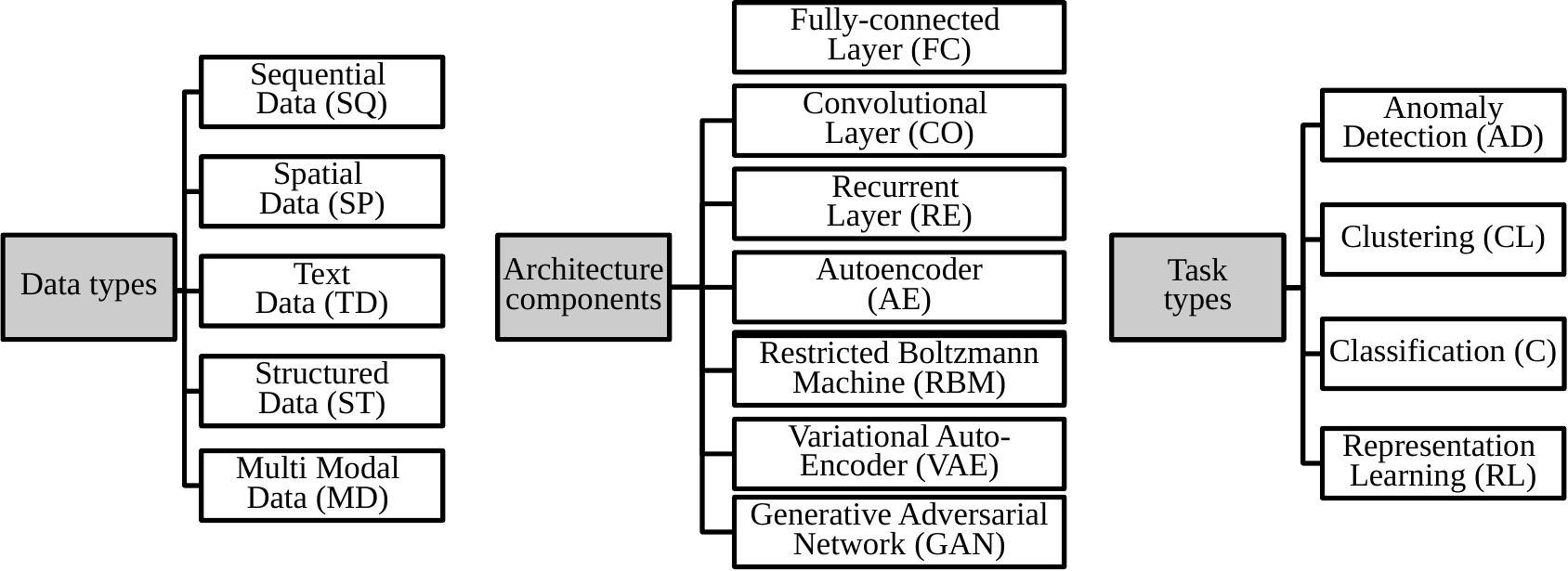}}
\caption{Dimensions of our categorization. }
\label{fig:catorization}
\end{center}
\end{figure}

Our classification consists of three dimensions: data-type, task-type, and model. Figure \ref{fig:cat-overview} depicts the high-level relationship between the three categorization dimensions of our survey. We use data to train a model to solve a task. Figure \ref{fig:catorization} depicts the abstract values of our categorization. 

We distinguish between the five data types that were introduced in Section~\ref{sec:bg:data-types}: spatial data (PD), sequential data (QD), structured data (ST), text data (TD) and multi-modal data. In addition to the abstract data type, we report on the concrete data type in the survey section. 

Furthermore, the next dimension of our classification is a distinction between four abstract task types: Classification (C), Representation Learning (RL), Anomaly detection (AD) and Clustering (CL). After this high-level, abstract categorization we group approaches together based on concrete security problems from different sub-domains of InfoSec. 

Finally, we distinguish which architecture components has been used to compose a model. The model choice depends largely on the data and the task that need to be solved, and multiple different architectures can be used to solve a task. We distinguish between fully connected layers (FC), convolutional layers (CO) which include pooling layers, recurrent layers (RE), AE (AE), restricted Boltzmann machines (RBM), variational AE (VAE) and generative adversarial networks (GAN). A model may contain more than one of these components. Often, model-architectures that were successful on a particular task get a name, such as LeNet-5 or AlexNet. When we found models that were heavily inspired by such architectures, we referred to that model. 

%
%
%
%
\section{Survey Methodology}
\label{sec:survey-methodology}

To obtain the literature for this survey, we followed the following protocol. 

We chose to restrict the sources of this survey by the venue. We selected security venues which published DL approaches and Machine Learning venues that published Security approaches that were about DL. Our venue selection criteria were the following. We selected the top 20 journals and conferences based on Security and Cryptography, Data mining, Artificial intelligence on Google Scholar's ranking. We added the top 20 Security conferences from Microsoft Academic Research. We searched for DL approaches in the Security conferences and Journals, and we searched for Security approaches, and DL approaches in the Machine learning areas. We limited the venues by name and the areas of DL and Security by keywords. We have obtained the keywords of DL by conventional technologies and the keywords for Security from the ACM CCS~\citep{ACM2012} and the IEEE taxonomy~\citep{IEEE2017}. 

To refine the selection of papers for this survey, we selected papers according to the following criteria. We excluded papers that only proposed ideas, but did not conduct any experiments. We excluded invited papers, talks, posters and workshop papers. We excluded papers that claimed to use DL, but only reference general resources such as Goodfellow et al.'s DL book~\citep{Goodfellow-et-al-2016-Book} without providing any additional detail that would enable researchers to reproduce results. Further, we excluded papers that were not explicitly aiming to achieve a security goal. For example, we excluded approaches on object detecting in videos, even though it could be argued that these approaches are useful for surveillance purposes.

We conducted the literature search on the 9th of May 2018 and we time-range of the survey was from 1st of January 2007 to the 8th of May 2018. The begin of 2007 was chosen because it approximately marks the begin of the advent of deep neural networks~\citep{schmidhuber2015deep}.

We used Scopus~\footnote{https://www.scopus.com/search/form.uri?display=basic} and Google Scholar~\footnote{https://scholar.google.com/}. We have conducted the keyword and venue based literature search on Scopus since it is the most comprehensive database of peer-reviewed computer sciences literature. For two venues, NDSS and Usenix-security, we used Google scholar because Scopus did not index them. 

Using the previous search criteria, we obtained 177 papers, which we reviewed more carefully. Of these papers, we only kept 77 that did fit our previously defined criteria. We list the complete list of venues that we included in our search in the Appendix \ref{app:venues}. 

%
%
%
%
\section{Deep Learning in Information Security}
\label{sec:survey}

Here, we review the papers that we found using the methodology of Section \ref{sec:survey-methodology}. On a high level we group the approaches into two parts. The first part consists of Sections \ref{sec:sequential-data} to \ref{sec:multi-modal-data} and presents approaches that apply DL algorithms to address security and privacy issues. The second part -- Section \ref{sec:sec-properties-dl-algorithms} presents approaches that address security and privacy issues of DL algorithms. 

We structured the first part based on the five data categories that we have defined in our classification scheme: sequential-, spatial-, structured-, text- and multi-modal data (see Section \ref{sec:classification-scheme}). For each of the data types, we group the reviewed approaches based on their task. For each of the tasks, we compare the different architectural decisions to solve the tasks and draw connections to advancements from in the DL community. After assessing applications of DL on these data types, we survey approaches that assess security properties of DL methods. We chose this data-type drive perspective of the use of DL in InfoSec because it reflects on the fact that the successful application of DL methods is data dependent, but not domain dependent. We believe that this perspective will help researches identify challenges and potential solutions to data of their domain more easily. 

As mentioned in \ref{sec:classification-scheme}, clearly distinguishing between different data types is often challenging. For example, an image, which is generally referred to as 2D spatial data, can also be viewed as a sequence of image patches and analyzed with methods that are appropriate for such data type. How to view the data and is ultimately a design choice which depends on the researcher. We group the approaches to the data types that are fed to the neural network, that is, if some form of feature processing changes the input data type, we use the processed version to categorize the approach.

%
%
%
%
\subsection{Sequential Data}
\label{sec:sequential-data}
\subsubsection{Classification on Sequential Data}
\paragraph{Function recognition}

Binary data or bytecode analysis is an important tool for malware analysis. One challenge of binary data analysis is function recognition, i.e., finding the start and end positions of software functions in a piece of code. Shin et al. tackle the challenge of function recognition from binary code by framing it as a binary classification task~\citep{shin2015recognizing}. They treat the binary code as a sequence of one-hot-encoded bytes and predict for each byte whether it is a start or end byte of a function. They evaluate a variety of sequential models such as RNNs, LSTMs~\citep{hochreiter1997long}, GRUs~\citep{chung2015gated}, and bi-directional RNNs~\citep{schuster1997bidirectional}, which are all well-suited for finding regularities in complex sequences. They train all their models using rmsprop~\citep{tieleman2012lecture}, which is a momentum-based back-propagation through time variant. They validate their approach on multiple datasets, and their best-performing model, a bi-directional RNN, achieves an F1-score between 98\% and 99\%, which means an average improvement of ~4\% to the then state-of-the-art, and the training and the evaluation are an order of magnitude faster than the state-of-the-art.

Instead of predicting start and stop bytes, Chua et al. aim to identify the function type given x86/x64 machine code~\citep{chua2017neural}. To this end, they treat the machine code as a sequence of instructions, which are embedded using neural word embeddings~\citep{bengio2003neural}. These embedded sequence instructions are then modeled using a 3-layer stacked GRU-RNN since the sequences are variable length and since the stateful nature of the recurrent neural network aids modeling sequence of instructions. The approach, which is named EKLAVYA, is evaluated on the machine code of two commonly used compilers: cc and clang and on multiple classification tasks. EKLAVYA achieves an average of up to 0.9748 accuracy for unoptimized binaries, and up to 0.839 accuracy for optimized code.

\paragraph{Traffic Classification}
Chen et al. propose a method for classification of mobile app traffic classification~\citep{Chen2017}. They capture the traffic, transform the headers into a 1-hot encoding and treat a sequence of packages as a sequence of vectors. They then train a 6-Layer custom CNN model to analyze the traffic. They validate their approach on a dataset captured from 20 apps with 355,235 requests and can achieve an average accuracy of 0.98. 

\paragraph{Encrypted Video Classification}
The MPEG-DASH streaming video standard allows to stream in an encrypted way. Schuster et al. show that it is possible to classify the videos based on the packet burst patterns in the video even though the videos are streamed in an encrypted way~\citep{schuster2017beauty}. A dash time series is created from captured TCP flows by aggregating the series into 0.25-second chunks. Then, they use a 1D CNN architecture to capture the locally correlated patterns that the encrypted sequences of burst sizes contain. The model uses three convolution layers to avoid an information bottleneck in the beginning. The convolutional layers are followed by a max-pooling layer and two dense layers. Dropout is used to prevent overfitting. The attack is validated on video datasets from different platforms, and in their best setting they achieve a recall of 0.988 with zero false positives.

\paragraph{Physical Side Channels}
Industrial control systems are controlled by programmable logic controllers (PLC). Han et al. propose ZEUS, a DL-based method for monitoring the integrity of the control flow of such a PLC~\citep{Han2017}. PLCs, when operate emit electromagnetic waves. Han et al. learn a program behavior model from spectrum sequences that are derived from execution traces. The execution traces are collected unobtrusively via a side channel by an inexpensive electromagnetic sensor. The spectrum sequences are modeled with stacked-LSTMs, which are well suited for modeling the sequential nature of the data. For validation, ZEUS is implemented on a commercial Allen Bradley PLC. ZEUS is capable of distinguishing between legitimate and malicious executions with an accuracy of 98.9\% while causing zero overhead to the PLC.

\paragraph{Mobile User Authentication}
Sun et al. use sequences of characters and the accelerator information to identify users on a mobile phone~\citep{Sun2017a}. They call their approach DeepService. To model the sequences they use a GRU~\citep{chung2014empirical}, which is a variant of the LSTM that uses fewer parameters, which is important for a mobile phone setting. DeepService is evaluated on a dataset that recorded 40 users, which was collected by Sun et al. On this dataset, DeepService achieves san F1-score of 0.93.

\paragraph{Steganalysis of Speech Data} 
Steganography is the discipline of hiding data in other data. Steganalysis attempts to discover and recover such hidden data. Lin et al. propose an RNN-based method to detect hidden data in raw audio data~\citep{Lin2018}. They segment speech data into small audio clips and frame the steganalysis problem as binary classification problem to differentiate between stego and cover. They train an LSTM-variant, called RNN-SM to classify on the raw data of the audio clips. Due to their stateful nature, LSTMs are well suited for modeling such data. RNN-SM is evaluated on two self-constructed datasets. One dataset contains 41 hours of Chinese speech data and the other contains 72 hours of English speech data. The accuracy depends on the language of the audio clips. In the best performing setting RNN-SM achieves an accuracy of up to 99.5\%.

%
%
\subsubsection{Representation Learning on Sequential Data}
\paragraph{Audio tampering detection } 
Adaptive multi-rate audio is a codec for compression of speech data that is commonly used in GSM networks. One way to manipulate such data is to decompress it to raw wave, modify the wave and re-compress it again. Luo et al. propose to extract features from compressed audio with a stacked under-complete AE SAE to identify whether a speech file has been recompressed or not~\citep{Luo2017}. To detect double-compression, the extracted feature sequences are classified with a universal background model, namely a Gaussian Mixture Model. The features are extracted by first splitting the audio file to frames, then normalizing them, then compress them via a SAE and finally fine-tune them with a binary classification layer. Two SAE models are trained, one for single compressed files and one for double compressed files. The under-completeness of the AE motivates the extraction of the most salient features, thereby enables to differ between the two compression types. Luo et al. validate their approach on the TIMIT database, where they achieve a 98\% accuracy.

%
%
\subsubsection{Clustering on Sequential Data}
\paragraph{Speech forensics}
Li et al. combine DL-based representation learning and spectral clustering to cluster mobile devices~\citep{Li2018}. They use an under-complete AE to compress speech files that originate from different mobile devices. This under-complete AE compresses the speech files and thereby extracts representations that capture the artifacts of each type of mobile device. The AE was pre-trained with restricted Boltzmann machines. Pre-trained AE have been showed to work well for modeling audio data~\citep{hinton2012deep}. The representations that are learned by the AE are then used to construct Gaussian super-vectors. Finally, these Gaussian super-vectors are clustered via a spectral clustering algorithm to determine the phone type of the speech recording. Li et al. evaluate their method on three datasets, MOBIPHONE, T-L-PHONE, and SCUTPHONE and in their best configuration their method achieves a maximum classification accuracy of 94.1\% among all possible label permutations.

{
\begin{table}
\caption{Classification using Deep Learning on sequential data in information security} 
\label{tbl:sequential-data}
\begin{center}
\begin{tabular}{lccc}
\hline %
\rule{0pt}{12pt} Data & Task & Architecture details & Approach \\
\hline %
\\
\multicolumn{4}{c}{ \bf Classification}\\[2pt]
\hline %
Binary Code	& Function Recognition & bi-RNN~\citep{schuster1997bidirectional} & \citep{shin2015recognizing} \\
Binary Code & Function Type Signatures & Word2Vec, GRU~\citep{mikolov2013efficient,chung2014empirical} & \citep{chua2017neural} \\
Speech Data	& Speaker Identification & bi-RNN~\citep{schuster1997bidirectional} & \citep{shin2015recognizing} \\ 
Encrypted Burst Sequences & Video Classification & 1D-CNN~\citep{schuster2017beauty} & \citep{schuster2017beauty} \\ 
HTTP Traffic & App classification & 6-layer CNN~\citep{Chen2017} & \citep{Chen2017} \\
Spectrum Sequences & Attack detection & stacked LSTM~\citep{gers1999learning}& \citep{Han2017} \\ 
Keystrokes & User authentication & GRU~\citep{chung2014empirical}& \citep{Sun2017a} \\
Speech data & Steganalysis & LSTM-variant & \citep{Lin2018} \\[5pt]
\multicolumn{4}{c}{ \bf Clustering}\\[2pt]
\hline %
Speech data & Speech forensics & Undercomplete AE & \citep{Li2018} \\[5pt]

\multicolumn{4}{c}{ \bf Representation Learning}\\[2pt]
\hline %
Speech data & Forging detection & Undercomplete SAE & \citep{Luo2017} \\
\end{tabular}
\end{center}
\end{table}
}

%
%
%
%
\subsection{Spatial data}
\label{sec:spatial-data}
%
%
\subsubsection{Classification on Spatial Data}

\paragraph{Breaking CAPTCHAs}
CAPTCHAs are a method for discriminating between automated and human agents~\citep{von2003captcha}. CAPTCHAS achieve this discrimination by posing an agent a challenge that is easy to solve for humans but hard for computers. An example of such a challenge is to ask whether an image contains a store or not, which is easy to solve for humans but not for computers. From their inception, CAPTCHAs have been continually evolving in an arms race with smarter bots that attempted to break them. DL algorithms proved to be very successful in analyzing images, especially distorted and incomplete images, which eventually led to attacks on CAPTCHAs using DL methods. 

Gao et al. attack hollow CAPTCHAs using a DL-based approach~\citep{Gao2013}. Hollow CAPTCHAs use contour lines to form connected characters. Their attack consists of two phases, a pre-processing phase, and a recognition phase. In the pre-processing phase, they repair the contour lines using Lee’s algorithm, fill the contour lines using a flooding algorithm, remove the noise component, and finally remove the contour line. The pre-processed data is then used to train a CNN-based model, which is trained to classify the characters on the picture by numbering and classifying the strokes and stroke combinations in the CAPTCHA. The final characters are determined by a depth-first tree search of a sequence that scored the highest predicted values. They selected a fully-connected version of LeNet-5's architecture~\citep{simard2003best}, which provides partial invariance to translation, rotation, scale, and deformation and has shown to work well on handwritten digits. They validate their attack on multiple datasets and their success-rate for attacks on hollow CPATHCA's ranges from 36\%-to 66\%. 

Algwil et al. use a similar CNN architecture and method to break Chinese CAPTCHAs~\citep{Algwil2016}. Chinese CAPTCHAs consist of distorted Chinese symbols which need to be recognized by the agent. Algwil et al. address this challenge with a multi-column CNN~\citep{ciregan2012multi}. A multi-column CNN is an ensemble of CNNs, whose predictions are averaged. The structure of each CNN is inspired by LeNet5, but it uses more and larger filter maps. Another difference is that the multi-column CNN uses max-pooling operations instead of trainable sub-sampling layers. This architecture also profits from the robustness of the CNNs to distortion, scaling, rotation and transformations. To attack the CAPTCHAs, they split each character into a set of radicals, i.e., composite elements, and classify input character to their radicals. Their best performing network is able to correctly solving a Chinese CAPTCHA challenge with 3755 categories with a test error rate of only 0.002\%.

reCAPTCHA is an image based CAPTCHA challenge developed by Google. Agents get a text-based assignment, and need to click all images that correspond to the assignment, for example: "Select all images with wine on it." Sivakorn et al. propose an automated reCAPTCHA breaker system~\citep{Sivakorn2016}. Their approach uses reverse image search to derive labels for images, word vectors~\citep{mikolov2013efficient} to derive similarity between assignment and a web-based on service that is based on deconvolutional neural networks~\citep{zeiler2011adaptive} to classify the images of the reCAPTCHA challenge. Deconvolutional neural networks use a form of sparse coding and latent switch variables that are computed for each image. These switch variables locally adapt the model's filters to learn a hierarchy of features for the input images, which enhance the classification performance on natural images. Their system is able to pass the reCAPTCHA challenge in 61.2\% of the attacks.

Finally, Osadachy et al. propose a CAPTCHA generation scheme, that is robust to DL-based CAPTCHA attacks~\citep{Osadchy2017a}. Their proposed CAPTCHA task is: given an adversarial image of one class, identify all images of the same class from a set of other images. An adversarial image is an image that has been modified with adversarial noise~\citep{szegedy2013intriguing}. Such adversarial noise prevents DL architectures from correctly classifying images, but the images appear to be visually similar for a human. They generated such adversarial images using a CNN-F architecture~\citep{chatfield2014return}. CNN-F is a variant of AlexNet~\citep{krizhevsky2012imagenet}, which was the winning architecture on the ImageNet LSVRC-2012 challenge. AlexNet has popularized three main techniques for image analysis: ReLUs as activation functions, Dropout instead of regularization to prevent overfitting and overlap pooling to reduce the network size. CNN-F is a stripped down version of Alex net that reduces the number of convolutional and fully-connected layers.   

\paragraph{Biometrics from 2D Spatial Data}
Biometrics are biological traits of a person which can be used to derive keys for identification in digital systems. Many biometrics are based on 2D spatial data, for example, images of fingerprints, faces or irises. One of the key challenges of creating biometrics from 2D spatial data is to be able to distinguish between subtle elements of the raw data that allow the identification. 

Faces are a standard way to identify people. One of the main challenges is that in real-life situations, images from faces will have distortions, different angles, and shades. Sun et al. proposed a DL-architecture that is well suited for such a task~\citep{Sun2014}. This architecture describes a CNN that consists of 4 convolutional layers, three max-pool layers and one fully connected layer, which yields an image representation. Additionally, they use local feature sharing in the third convolutional layer to encourage more diverse high-level features in different regions. Finally, the outputs of the last max-pool layer and the output of the last convolutional layer are shared inputs to the fully connected layer. This connection prevents an information bottleneck caused by the last convolutional layer and is essential for learning good representations. Their approach manages to achieve a 99.15\% accuracy on the LFW dataset with 5749 different identities when their network was augmented with additional, external training data. Zhong et al. use a similar approach for the task of face-attribute classification~\citep{Zhong2016a}. They compare VGGnet~\citep{simonyan2014very} with FaceNet~\citep{schroff2015facenet} and conclude that both are suitable to extract facial attributes. 
Instead of Iris as a biometric, Zhao et al. focus on the region around the eyes -- the periocular region -- for authentication~\citep{Zhao2017}. They use AlexNet to classify the eyes and derive semantic information such as age, or gender using additional CNNs that are trained on the surrounding regions of the eye. Goswami et al. use a combination of an SDAE and a DBM to extract features from multiple face images which they extract from video clips and then use an MLP to authenticate people via their faces~\citep{Goswami2017}. They pre-filter the frames with a high information content as input for the classification network. At a false-accept rate of 0.01, they achieve up to 97\% accuracy when validated against the point and shoot challenge database, and 95\% against the Youtube faces dataset. 

Instead of directly classifying faces with a CNN for biometrics, Liu et al. propose to learn a metric space for faces~\citep{Liu2016b}. Their approach is inspired by triplet networks of Schroff et al. ~\citep{schroff2015facenet}. Triplet networks learn a metric space by solving a ranking problem between three images that are represented by the same CNN. The parameters of such a model are learned by minimizing a triplet loss. Liu et al. pre-train their model on the CASIA Web-Face Dataset and evaluate CASIA NIR-VIS 2.0 face dataset, where they achieve a rank one accuracy of 95.74\%. The second best approach achieved an accuracy of 86.16\%.

Fingerprints are another standard way to identify people. One important feature to match fingerprint images is the local orientation field. Cao et al. frame the local orientation field extraction as a DL-based classification task. Given a fingerprint image, they predict the pixel probabilities of this image that a pixel belongs to a local orientation field. They base their CNN model on LeNet-5, but adjust the filter size and extend the model by dropout and local response normalization~\citep{krizhevsky2012imagenet}. In addition to that, they augment their training data with texture noise. They evaluate their method on on the NIST SD27 dataset, where they achieve a pixel-class root mean square deviation of 13.51, which presents an improvement for all latents of 7.36\%. 

Segmentation is the computer vision task of splitting an image into multiple parts so that they can be processed better. In biometrics, iris segmentation aims to separate the parts of an image that belongs to an iris from the remaining images, so that the iris can be used for biometric identification. Liu et al. propose a DL-based approach for biometric iris segmentation~\citep{Liu2016a}. They use an ensemble of pre-trained VGG19-net~\citep{simonyan2014very}, which is fine-tuned on two iris datasets ( UBIRIS.v2, CASIA.v4), to perform a pixel-based, binary classification task. VGG-net~\citep{simonyan2014very} advanced the state-of-the-art by two novelties: they increased the depth of the network by using many layers of smaller, 3x3 filters and they used 1x1 convolution, to increase the non-linearity of the decision function without reducing the size of the receptive fields. Simonyan et al. evaluate their approach on the UBRIRIS.v2 and the CASIA.v4 datasets and achieved a pixel segmentation error rate of less than 1\%. Qin and Yacoubi use a similar approach for finger-vein segmentation. However, they use a custom CNN instead of a pre-trained VGG19-net~\citep{Qin2017}. Their approach is validated on two finger-vein databases and achieves an equal-error-rate of 1.69 in the best setting. 

Peoples' appearances are a form of soft-biometrics that can be used to identify people in surveillance videos. Zhu et al. uses multiple AlexNet-based CNNs to classify properties of people from surveillance images to identify them~\citep{Zhu2015}. They use an AlexNet for a binary classification task for each of the desired attributes. They validate their approach on the GRID and VIPeR databases, and their method out-performed the baseline, an SVM based classifier, by between 6\% and 10\% percent of accuracy on average. 
Hand and Chellappa also classify attributes of faces. They frame the problem as multi-class approach and design a particular architecture for the task: MCNN-AUX~\citep{Hand20174068}. The MCNN-AUX consists of two shared convolution-max pool layers for all attributes and one CO layer followed by two FC layers for each task. Additionally, the data of all attributes are used for classification. Gonzalez-Sosa et al. compare a VGG-net variant with commercial off the shelf face attribute estimators, and find that soft biometrics improve the verification accuracy of hard biometric matchers~\citep{Gonzalez-Sosa2018}. 
Another soft biometric is gait, i.e., how people walk. Shiraga et al. use a CNNs called GEINet to classify peoples gait~\citep{Shiraga2016a}. The input to GEINet is gait energy images (GEI), a particular type of images that are extracted from multiple frames of images that show people's walk. GEINet is based in AlexNet~\citep{krizhevsky2012imagenet}. Shirage et al. reduce the depth of the network because of the different nature of the task, that is to classify GEI the network needs to focus on subtle inter-subject images. During the evaluation, GEINet consistently manages to outperform baseline approaches by a significant margin; In their best experiment settings, GEINet achieves a rank-1 identification rate of 94.6\%. Age, gender, and ethnicity are other common soft-biometrics. Narang et al. empirically validate that VGGnet is suitable for age, gender, and ethnicity classification~\citep{Narang2016a}.
Azimpourkivi et al. propose another soft-biometric for authentication for mobile-phone pictures of objects~\citep{Azimpourkivi2017}. They use an InceptionV3 CNN~ \citep{Azimpourkivi2017} to derive features from pictures a mobile phone user takes. The Inception architecture introduced inception modules, which consist of multiple convolutional filters of different sizes. These filters are concatenated, and allow the network to decided which filter size it needs at a given layer in the network to solve the task at hand. To construct images hashes that are not sensitive to subtle changes, they use a particular version of locality sensitive hashing which allows them to group similar image features next to each other. Azimpourkivi et al. demonstrate that their approach is robust to real image-, synthetic image-, synthetic credential- and object guessing attacks, and show, that the Shannon entropy of their soft biometric is higher than that of fingerprints. Additionally, unlike real biometrics, soft biometrics that are derived from object pictures can be easily exchanged.

People's voices are also biometric data that can be used to identify a person. Generally, audio data is presented in waveform, which is a form of sequential data. However, a common pre-processing step in speech recognition is to calculate the Mel-frequency cepstral coefficients (MFCC), which represents the speech waves as 2D spatial data. Uzan and Wolf combines MFCCs and an AlexNet-based CNN~\citep{krizhevsky2012imagenet} to identify speakers~\citep{Uzan2015}. They validate their approach on their own dataset with 101 persons and 4584 utterances and their method achieves an utterance classification accuracy of 63.5\%.

Another challenge in biometrics is to determine whether a biometric is authentic or not, i.e., to detect it has been tampered with or not. Menotti et al. propose a DL-based framework for detecting spoofed biometrics~\citep{Menotti2015}. At the heart of their framework is a CNN that performs a binary classification task on images such as iris scans, fingerprint scans or face images. The goal of this task is to distinguish between fake and real biometric images. They evaluate two different strategies, architecture optimization, and filter optimization. In their architecture optimization strategy, they leave the parameters random but change the architecture to fit the problem. In their filter optimization, they fix the architecture but try to find the optimal hyper parameter settings. To do so, they create spoofnet. Spoofnet is a CNN, which is an architecture that is based on Cuda-net~\citep{krizhevsky2012cuda}. Spoofnet is designed to handle subtle changes in data better because of the modified filter sizes. In contrast to their architecture optimization strategy, they train the parameters of spoofnet with SGD. They validate their approach on seven different databases and achieved excellent results compared to the state-of-the-art. 
Nougeria et al. use pre-trained VGG-19, which is fine-tuned on fingerprints, for distinguishing between fake and live fingerprints~\citep{Nogueira2016}. Their best performing model achieves a 97.1\% test-accuracy. Instead of CNNs, Bharati et al. use supervised deep Boltzmann Machines~\citep{Bharati2016} to detect the spoofing of face images. Their method outperformes an SVM on their custom datasets. Conceptually very similar to spoofing detection is biometric liveness detection. 

\paragraph{Detecting Privacy Infringing Material}
Pictures may contain privacy-sensitive information, such as drivers license or other confidential material. Tran et al. propose a CNN-based method to detect such material~\citep{Tran2016}. They combine a modified version of AlexNet~ \citep{krizhevsky2012imagenet} with a CNN for sentiment analysis~\citep{you2015robust} to detect such material. They name this model PCNH. They use 200 classes from the ImageNet challenge~\citep{russakovsky2015imagenet} to pre-train PCNH, and then train PCNH on categories of sensitive material. They validate their approach on their own and a public dataset with an F1-score of 0.85 and 0.89, respectively. Yu et al. tackle a similar challenge with a hierarchical CNN~\citep{Yu2017}. They combine deep CNN for object detection with a tree-based classifier over the visual tree. The object identification focuses on detecting infringing privacy object, and the tree-based classifier combines them. 

\paragraph{Steganalysis and Steganography}
Steganalysis is the process of analyzing data to find out whether secrets have been hidden in the data or not. One way to hide information is hiding it in images. One of the key challenge of detection such hidden data requires to identify subtle changes in the data. Ye et al. propose a DL-based method for image steganalysis, where they use a deep neural network to detect whether information has been hidden or not~\citep{Ye2017}. Their proposed CNN architecture has an adaption to suit the data for the task at hand. First, the first three layers are convolutional layers only, which is important not to lose any information. Also, no Dropout or normalization is used. Furthermore, instead of Max-Pooling layers, they use mean pooling layers. In total, their network is ten layers deep. Apart from that, they use a truncated linear unit (TLU) activation function in the first layer, which reflects that most steganographic hidden data is between -1 and 1. The weights in the first convolutional layer are initialized with high-pass filters, which helps the CNN to ignore the image content and instead focuses on the hidden information. For training their network, Ye et al. adopt a curriculum learning scheme, which selects the order of the images that are used for training. This ordering is essential for detecting information that has been hidden with a meager bpp rate. They evaluate their method on multiple steganographic schemes and bpp rates, and their approach consistently outperforms the baseline approaches. Their detection error rate ranged from 0.46 to 0.14. 
Zeng et al. propose to use the residuals and an ensemble of CNNs to detect steganography in images~\citep{Zeng2018}. As input to their network, they use quantized, truncated residuals, with three different parameters for the quantization and truncation. The three different input sets are used to train three different CNNs. The CNN architecture is based on insights of the steganalysis network of Xu et al.~\citep{xu2016structural}, but they use ReLUs instead of TanH. Zeng et al. validate their approach on ImageNet and three steganographic algorithms, and their best performing model achieves a detection accuracy of 74.5\%. 

\paragraph{Authenticity of Goods and Origin Determination}
Sharma et al. applied AlexNet on the problem of separating real from counterfeited products~\citep{Sharma2017}. To detect faked material such as fake leather, they use microscopical images to build a model of the texture of the objects. Then, they use AlexNet to classify different types of material. Depending on the material, the proposed method reaches a test accuracy of up to 98.3\%. 
Similarly, Ferreira proposes a DL-based method to determine the device that a printed document was printed with~\citep{Ferreira2017}. Printed documents differ from each other in very subtle ways, depending on the manufacture of that printer. To find these subtle differences, Ferreira et al. use a CNN architecture with two convolutions and two sub-sampling layers. They use three different represent the input in three different ways, raw pixel, median residual filter and average residual filter. These different input representations were either combined with an early or a late fusion strategy. Early fusion combines multiple representations of the same data into one input data vector, and late fusion combines the output of multiple CNNs. Ferreira et al. evaluated their approach on a dataset which consists of 120 documents and ten printers. Using the best overall settings, their method was able to attribute printers with an accuracy of 97.33\%.

\paragraph{Forensic Image Analysis}
A problem law enforcement faces when searching suspects computer for illegal pornographic material is the vast amount of pictures that can be stored on a computer, many of which are irrelevant. In their work, Mayer et al. evaluate DL-based image classification services on their capabilities to identify and pornographic material from non-pornographic material~\citep{Mayer2017a}. They evaluate Yahoo's service, Illustration2Vec, and Clarifai, NudeDetect and Sightengine. They find, that when such services are used to rank such images on their not safe for work character, that, on average, relevant images are discovered on positions 8 and nine instead of position 1,463.

\paragraph{Watermarking}
Watermarks address the problem of data attribution. For example, watermarks are added to copyright protected material to identify the owner this material. Kandi et al. propose to use a convolutional neural AE to add an invisible watermark to image~\citep{Kandi2017}. A CNN AE is an AE that has CNN as encoding and decoding networks.

\paragraph{Detecting Polymorphic Malware}
Nguyen et al. propose a method for detecting polymorphic malware from binary files via a CNN~\citep{Nguyen2018}. To be able to use CNNs, they first extract the control flow graph from the binary. They transform the graph into an adjacency matrix, where a vertex of the control flow graph is considered as a state, and three values describe each state: register, flag, and memory. The register, flag, and memory are mapped to the color of red, green and blue pixels. This representation allows describing similarity in a program state. Even if two variants of the same malware have different execution codes, the core malicious actions retain in the same are of such an image. The corresponding image will differ, and CNNs are highly suitable for detecting similarities between such images with losing spatial correlation. Nguyen et al. evaluate different architectures on a set of polymorphic malware, and they find that the YOLO-architecture~\citep{redmon2016you} is the best performing. The YOLO-architecture can detect polymorphic malware with an accuracy of up to 97.69\%.

\paragraph{Detecting Website De-facement}
Website defacements are unauthorized changes to a website that can lead to loss of reputation or financial gain. Borgolte et al. use deep neural networks for detecting web site-defacements from website screenshots~\citep{borgolte2015meerkat}. They train their model on screenshots from web-pages, which they automatically collected using a web-crawler. Since the screenshots are comparably large, they extract a window uniformly sampled from the center of the webpage. This extracted window is then compressed with a stacked denoising AE., and finally is classified to normal or defaced using an AlexNet-like network. They validate their approach on their own, large-scale defacement dataset and their method achieves a true positive rate of up to 98.81\%.

{
\small
\begin{table}
\caption{Classification using Deep Learning on spatial data in information security}
\label{tbl:sequential-data-overview}
\begin{center}
\begin{tabular}{lccc}
\rule{0pt}{12pt} Data & Task & Architecture details & Approach \\[2pt]
\hline %
Images	& Breaking hollow CAPTCHAS & LeNet-5 variant~\citep{simard2003best} 	 & \citep{Gao2013} \\ 
Images	& Breaking Chinese CAPTCHAS & ensemble of LeNet-5 variants~\citep{ciregan2012multi} & \citep{Algwil2016} \\ 
Images	& Breaking reCAPTCHAS & adaptive deconvolutional Net~\citep{zeiler2011adaptive} & \citep{Sivakorn2016} \\ 
Images & Generating CAPTCHAS & AlexNet variant~\citep{chatfield2014return} & \citep{Osadchy2017a}  \\
Images 	& Face verification & DeepID2~\citep{Sun2014} & \citep{Sun2014} \\
Images & Face verification & SDAE, DBM & \citep{Goswami2017} \\ 
Images 	& Fingerprint orientation field extraction & LeNet-5 variant~\citep{lecun1998gradient} & \citep{Cao2015} \\

Images	& Spoofing detection & Spoofnet~\citep{Menotti2015}, variant of cuda-convnet~\citep{krizhevsky2012cuda} & \citep{Menotti2015} \\
Images	& Spoofing detection & DBM~\citep{salakhutdinov2010efficient} & \citep{Bharati2016} \\
Images	& Liveness detection & VGG19~\citep{simonyan2014very} & \citep{Nogueira2016} \\
Images	& Person identification & multiple AlexNet~\citep{krizhevsky2012imagenet} & \citep{Zhu2015} \\
Images & Gait classification & GEINet, a reduced AlexNet~\citep{krizhevsky2012imagenet} & \citep{Shiraga2016a} \\
Images & Soft biometrics classification & VGG19~\citep{simonyan2014very} & \citep{Narang2016a} \\
Images & Soft biometrics classification & VGGNet~variant~\citep{parkhi2015deep} & \citep{Gonzalez-Sosa2018} \\
Images & Face verfication & FaceNet~\citep{schroff2015facenet} \\
Images	& Iris segmentation & VGG19~\citep{simonyan2014very} & \citep{Liu2016a} \\
Images & Iris verification & AlexNet, custom CNN & \citep{Zhao2017} \\ 
Images & Iris verification & AlexNet-variant & \citep{Proenca2018} \\
Images	& Privacy infringing material & AlexNet variant~\citep{krizhevsky2012imagenet}, CNN~\citep{you2015robust} & \citep{Tran2016} \\
Images & Privacy infringing material & Custom hierarchical CNN inspired by~\citep{krizhevsky2012imagenet,donahue2014decaf} & \citep{Yu2017} \\
Images	& Steganalysis & custom CNN~\citep{Ye2017} & \citep{Ye2017}\\ 
Images & Steganalysis & custom CNN based on~\citep{xu2016structural} & \citep{Zeng2018} \\
Images & Alternative Biometrics & InceptionV3~\citep{szegedy2016rethinking} & \citep{Azimpourkivi2017} \\
Images	& Counterfeit detection & AlexNet-variant~\citep{krizhevsky2012imagenet} & \citep{Sharma2017} \\
Images	& Forensic image analysis & DL based services & \citep{Mayer2017a} \\
Images & Printer attribution & Custom CNN~\citep{Ferreira2017} & \citep{Ferreira2017} \\ 
Images & Defacement detection & Denoising AutoEncoder, AlexNet~\citep{krizhevsky2012imagenet} & \citep{borgolte2015meerkat} \\
CFG Graphs & Malware detection & Yolo~\citep{redmon2016you}, LeNet-5~\citep{lecun1998gradient} & \citep{Nguyen2018} \\
Voice	& speaker identification & AlexNet~\citep{krizhevsky2012imagenet} & \citep{Uzan2015} \\
\end{tabular}

\end{center}
\end{table}
}

%
%
\subsubsection{Representation Learning on Spatial Data}

Forensic face-sketches are pictures of faces of crime suspects that are hand-drawn by an artist with help from instructions of eye-witnesses. Mittal et al. propose a DL-based approach to match forensic face-sketches with images of people~\citep{Mittal2015}. Their approach consists of two phases. First, they learn representations from face images in an unsupervised way. Their model for representation learning is a combination of stacked denoising AE~\citep{vincent2008extracting} and deep belief networks~\citep{hinton2009deep}. They first train this model on real people's faces and fine-tune it on the face sketches. Secondly, they train an MLP in a directed way to predict a match-score between pairs of images. The input to this MLP is the concatenated learned representation for the real image and the sketch image. They evaluate their approach on multiple databases achieved an accuracy score of 60.2\%, which is a significant improvement to the second best approach, which matched only 50.7\% face-sketch pairs correctly. Galea and Farrugia tackle the same task of matching with a combination of VGGnet and a triplet network, which is called DEEPS~\citep{Galea2018}. They validate DEEPS on the UoM-SGFS 3D sketch database and achieved a Rank-1 matching rate of 52.17\%.

Wang et al. use the representation-learning capabilities of neural networks to increase the performance of Commercial Off-the-Shelf (COTS) forensic face matchers~\citep{Wang2015}. They use a pre-trained AlexNet~\citep{krizhevsky2012imagenet} to derive a vector representation of face images of the last fully connected layer before the classification layer. They use cosine similarity as a distance metric between faces and rank the images using probabilistic hypergraph ranking~\citep{huang2010image}. Then, they select the top $k$ images as an input to the COTS PittPatt matcher, where $k$ is a hyperparameter. They validate their approach on a combination of three databases, the LFW, WLFDB and PSCO database and their approach consistently scores a higher mean average precision than the PittPatt matcher alone. 

\paragraph{Spoofing Detection}
Czajka compares hand-crafted features to learned features for suitability on the task of detecting a spoofing attack with rotated biometrics~\citep{Czajka2017}. The final classification is conducted with an SVM in both cases. They evaluated their approach on multiple datasets from different sensors and found that the hand-crafted features performed better in the cross-sensor test and the learned features performed better in the same-sensor test.

\paragraph{Mobile Iris Verification}
Zhang et al. explore the suitability of learned representations for biometric iris verification on a mobile phone~\citep{Zhang2016a}. They use a model that is based on the ideas of Zagorykyo et al., who have proposed to compare two image patches by a 2-channel CNN that takes as input a pair of images for each channel~\citep{zagoruyko2015learning}. This model has one output and is trained on a hinge-loss to regress on the similarity between the two patches. Zhang et al. combine two types of representations for their iris-verification approach: the one derived by a 2-channel CNN and representations derived by optimized ordinal measures. They demonstrate that the FER rate for iris verification can be significantly decreased by combining these two features.

\paragraph{Face Verification}
Gao et al. address the task of face identification by using a supervised deep, denoising stacked AE. This model learns face representations that are robust to differences such as illumination, expression or occlusion~\citep{Gao2015}. Their architecture consists of a two-layer denoising AE. The input was a canonical face image of a person, i.e., a frontal face image with a neutral expression and normal illumination, and the "noisy" input to reconstruct were images of the same person. They use their AE to extract face features, and sparse representation-based classification for the face verification task. On the LFW dataset, they achieve a mean classification accuracy of 85.48\%. Bharadwaj et al. applied a similar approach on baby faces~\citep{Bharadwaj2016}. They use stacked denoising AE to learn robust representations of baby-faces, and a one-shot, single-class support vector machine to classify the baby faces. Their system achieves a verification accuracy of 63.4\% with a false accept rate of 0.1\%.
Noroozi et al. propose to learn representations for biometrics with an architecture called SEVEN~\citep{Noroozi2017}. SEVEN combines a convolutional AE~\citep{masci2011stacked} with a Siamese network. The combination of these components allows learning general salient and discriminative features of the data. Also, it enables the network to be trained in an end-to-end fashion. 

{
\begin{table}
\caption{Representation Learning using Deep Learning on spatial data in information security}
\label{tbl:spatial-data}
\begin{center}
\begin{tabular}{lccc}
\rule{0pt}{12pt} Data & Task & Architecture details & Approach \\[2pt]
\hline %
Images	& Forensic sketch matching & SDAE~\citep{vincent2008extracting}, DBN~\citep{hinton2009deep} & \citep{Mittal2015} \\
Images & Forensic sketch matching & FaceNet~\citep{schroff2015facenet}, VGGNet & \citep{Galea2018} \\
Images	& Face filtering & AlexNet~\citep{krizhevsky2012imagenet} & \citep{Wang2015} \\
Images 	& Face verification & SDAE~\citep{vincent2008extracting} & \citep{Gao2015} \\
Images 	& Baby face verification & SDAE~\citep{vincent2008extracting} & \citep{Bharadwaj2016} \\
Images	& Spoofing detection & CNN~\citep{Czajka2017} & \citep{Czajka2017} \\
Images	& Mobile iris verification & 2Channel CNN~\citep{zagoruyko2015learning} & \citep{Zhang2016a}\\
Images & RL for biometrics & Siamese Network, CNN AE & \citep{Noroozi2017} \\
\end{tabular}

\end{center}
\end{table}
}

%
%
%
%
\subsection{Structured data}
\label{sec:structured-data}
\subsubsection{Classification on Structured Data}

Alsulami et al. propose to identify the authors of source code by features that are extracted by an LSTM from the source code~\citep{Alsulami2017}. To achieve this, they first generate the abstract syntax tree (AST) from the source code. Then, this AST is traversed depth-first to create a sequence of nodes. Each node is embedded via an embedding layer, and a model is learned from such sequences of nodes via a bi-directional LSTM. Alsulami at al. evaluate their method via code obtained from public source code repositories, and achieve an author classification accuracy of up to 96\%. 

\paragraph{DL-based PUF Verification} 
Physically Unclonable Functions (PUFs) can be used in an authentication system in a challenge-response based protocol. Yashiro et al. propose DAPUF, a strong, arbiter PUF-based system for authentication of chips, that can be used to authenticate them. They show that their system is resistant to DL based attacks~\citep{Yashiro2016}. To carry out these attacks, they use a stacked denoising AE followed by a FC layer to differentiate between fake and genuine PUFs in a binary classification task.

\paragraph{Network Intrusion Detection}
Network intrusion detection is the task of analyzing network traffic data or data from other components of a network to identify malign actions. 
Osada et al. propose a network intrusion detection method that uses latent representations of network traffic to identify malign actions~\citep{Osada2017}. Osada et al. propose a semi-supervised version of VAEs, the forked variational AE to address this task. The forked VAE learns a representation from the traffic, and subsequently the mean of the latent space defined by the VAE is used as input to a classifier that , and predicts whether an example is benign or not. The error is back-propagated and combined with the VAE reconstruction error. Details on feature extraction and how to overcome problems of training a discrete VAE are omitted. Osada et al. evaluate their approach on the NSL-KDD and Kyoto2006+ datasets. Adding 100 labeled examples increased the absolute recall-rate by 4.4\% points the total false-positive rate by 0.019\%. 
Similarly, Aminanto et al. propose to learn features from network data to detect impersonation attacks~\citep{8067440}. They extract the features using a sparse deep AE from the existing network data features, i.e., they compress the already complex features. They use this learned frame representation to learn a feed-forward MLP. They evaluate their approach on the Aegean Wireless Network Intrusion Detection Dataset (AWID) and achieve a per-frame classification accuracy of 99.91\% and a false positive rate of 0.012. 

\paragraph{Drive-by Attack Detection}
Drive-by attacks are attacks that infect clients that visit a web page by exploiting client-side vulnerabilities. Shibanara et al. propose a method for detecting such attacks by classifying the sequences of URLs into benign and malicious sequences~\citep{Shibahara2017a}. They extract 17 features from each of the URLs that are loaded when a client visits a web page, and classify these sequences using a CNN. Shibanara et al. propose an Event Denoising CNN (EDCNN) to suit the data at hand. This EDCNN has an allocation layer, which rearranges the values of the input layer to convolve over two URLs whose order is similar. Additionally, they use a spatial pooling layer to summarize sequences of different length~\citep{he2014spatial}. They evaluate their approach on a data set that they collected using a honey client and which consists of 17,877 malicious and 41,127 benign URL sequences. On this dataset, an EDCNN achieves a false-positive rate of 0.148 and an F1-score of 0.72, which outperforms a regular CNN, which achieves a false-positive rate of 0.276 and an F1-score of 0.59.

\subsubsection{Representation Learning on Structured Data }
\paragraph{Cross-Platform Binary Code Similarity Detection}
Cross-Platform binary code similarity detection aims to identify whether two pieces of binary function code from different platforms describe the same function. To address this problem, Xu et al. propose to learn a metric space for functions~\citep{Xu2017a}. They do so by combining a Siamese network~\citep{bromley1994signature} with a Structure2Vec model~\citep{dai2016discriminative}. That is, they treat the function code as a graph, hence chose Structure2Vec as graph embedding network in a Siamese learning setting. Xu et al. demonstrate on four datasets that their approach outperforms state-of-the-art approaches by large margins. Additionally, they show that the embedding was learned 20x faster and that the embedding maps binary codes of the same function close to each other, and functions from different functions further apart. 

{
\small
\begin{table}
\caption{Deep Learning in information security on structured data} 
\label{tbl:structured-data}
\begin{center}
\begin{tabular}{lccc}
\hline %
\rule{0pt}{12pt} Data & Task & Architecture details & Approach \\[2pt]
\hline %

\\
\multicolumn{4}{c}{ \bf Classification}\\[2pt]
\hline %

Abstract Syntax Tree & Author attribution & bi-LSTM~\citep{schuster1997bidirectional,hochreiter1997long} &\citep{Alsulami2017}  \\ 
Wireless Network traffic & Intrusion Detection & AE, MLP & ~\citep{8067440} \\
Challenge and response pairs & PUF verification & SDAE~\citep{vincent2008extracting}, RBM & \citep{Yashiro2016} \\
Network traffic & Intrusion Detection & VAE-variant~\citep{kingma2013auto} & ~\citep{Osada2017} \\
URL-Sequences 	& Drive-by-attack detection & EDCNNs~\citep{Shibahara2017a} & ~\citep{Shibahara2017a} \\ [5pt]
\hline
\\
\multicolumn{4}{c}{ \bf Representation Learning}\\[2pt]
\hline
Execution Graph & Binary Code Embedding & Structure2Vec~\citep{dai2016discriminative}, Siamese Network~\citep{bromley1994signature} &\citep{Xu2017a}  \\ 
\end{tabular}
\end{center}
\end{table}
}

\subsection{Text data}
\label{sec:text-data}
\subsubsection{Classification on Text Data}
\paragraph{Password guessing attacks}
Melicher et al. propose DL-based password guessing attacks, which they use to measure the strength of passwords~\citep{melicher2016fast}. Their idea is inspired by works of Sutskever et al., who have successfully demonstrated that RNNs could be used to build language models for text, which in turn could also be used for generating text. Melicher et al. use fine-tuned LSTMs~\citep{jozefowicz2015empirical} to learn a model for the characters of passwords. This model can be used to predict the likelihood of a given password, and given the likelihood, they calculate the strength. They train their model on a large list of publicly available passwords and show, that a DL-based password model performs better at predicting than Markov chains or Context-Free Grammars.

\paragraph{Creating and Detecting Fake Reviews}
E-commerce sites sell products which partly derive their reputation via user reviews. Yao et al. use a DL-based language model to create fake reviews as well as a defensive model~\citep{Yao2017}. Their fake review language model is a word-based language model based on LSTMs~\citep{hochreiter1997long}. LSTMs have been shown to model statistical properties of texts well~\citep{cho2014learning,sutskever2014sequence, bahdanau2014neural}). They use a large text corpus, the Yelp review database for training, and fine-tune the generated reviews by a simple noun replacement strategy. This noun replacement strategy refines the review to the context of the review. Yao et al. experimentally show, that about 96\% of the fake reviews pass automated plagiarism detection, and human recall and precision are 0.160 and 0.406. To defend against such automated attacks, Yao et al. propose to use a character-based language model, as they find that generated texts have a statistically detectable difference in character distribution to text written by real people. Their defense achieves a prevision of 0.98 and a recall of 0.97. 

\subsubsection{Anomaly Detection on Text Data}
\paragraph{Log analysis}
System logs keep track of activities happening on a computing system. In case of system failures or attacks, system logs can be used to debug such failures or reveal knowledge about an attack. Du et al. propose a neural sequence model-based approach for anomaly detection in system logs called DeepLog~\citep{Du2017}. Their neural sequence model consists of a stacked LSTM~\citep{hochreiter1997long}. To build the sequence model, they manually parse lines to a specific event IDs to construct sequences of events. Such sequences describe workflows on the analyzed system. The normal workflow model is built from a set of workflows. The model continuously tries to predict next event IDs, and if they predicted event ID is above a certain mean squared error for the actual error of the event, the system raises an alarm. Du et al. integrate user feedback to update the model if a false positive has been detected. DeepLog is evaluated on two large system logs, and it achieves an F1-score of up to 96\%.

\subsubsection{Clustering on Text Data}
\paragraph{Log Analysis}
Thaler et al. propose an unsupervised method for signature extraction from forensic logs~\citep{Thaler2017b}. This approach groups log lines based on the print statement that originated their log lines. Their proposed method combines a Neural language model~\citep{bahdanau2014neural} and with a clustering method. They train the neural language model as an RNN-autoencoder to learn a representation of the loglines, which captures the complex dependency between mutable and immutable parts of a logline. These representations are then clustered. They evaluate their approach on three datasets, one self-created and two public system logs with 11,023, 474,796 and 716,577 log lines. Their method clusters log lines with a V-measure of 0.930 to 0.948. 

{
\small
\begin{table}
\caption{DL in InfoSec on text data} 
\label{tbl:text-data}
\begin{center}
\begin{tabular}{lccc}

\hline
\rule{0pt}{12pt} Data & Task & Architecture details & Approach \\[2pt]
\hline %

\\
\multicolumn{4}{c}{ \bf Classification}\\[2pt]
\hline

Characters & Password strength guessing & LSTM~\citep{jozefowicz2015empirical} & \citep{melicher2016fast} \\ 
Text	& Detecting fake reviews & LSTM~\citep{hochreiter1997long} & \citep{Yao2017} \\ 

\\
\multicolumn{4}{c}{ \bf Anomaly Detection}\\[2pt]
\hline
Logs	& Anomaly detection & LSTM~\citep{hochreiter1997long} & \citep{Du2017} \\ 
\\
\multicolumn{4}{c}{ \bf Clustering}\\[2pt]
\hline
Logs	& Signature Extraction & Neural Language Model~\citep{bahdanau2014neural} & \citep{Thaler2017b} \\ 
\end{tabular}
\end{center}
\end{table}
}

%
%
%
%
\subsection{Multi-modal data}
\label{sec:multi-modal-data}

%
%
\subsubsection{Classification on Multi-Modal Data}
\paragraph{Cyber bullying detection}
Zhong et al. propose a DL-based method to detect bullying in cyber space~\citep{Zhong2016b}. To do so, they build a model from images and comments of Instagram. For classification of the images, they use a pre-trained version of AlexNet, which they fine-tuned using their dataset. For representing text, they used word2vec~\citep{NIPS2013_5021} and other, shallow representations such as bag-of-words. Zhong et al. experimented with different configurations and combinations and conclude, that the title of the post is one of the strongest predictors of cyber bullying.

\paragraph{ECG Biometrics}
With cheap mobile sensors available, ECG can be used as a modality for biometrics. Da Silva Luz et al. propose a method for ECG-based biometrics, which treats the ECG data as both, raw sequential data and spatial data~\citep{DaSilvaLuz2018}. They derive the 2D spatial data from the raw ECG data by calculating a spectrogram. They build two models from these two data modalities, a 1D, and a 2D CNN. The outputs of these two networks are merged using either a sum, a multiplication or a mean rule. To capture a whole ECG cycle, both CNN models have large first layer convolutions. The method is evaluated on multiple datasets. The fusion of the two model results consistently increases the accuracy of the model.

{
\small
\begin{table}
\caption{Deep Learning in information security on multi-modal data} 
\label{tbl:multi-modal-data}
\begin{center}
\begin{tabular}{lccc}
\hline %
\rule{0pt}{12pt} Data & Task & Architecture details & Approach \\[2pt]
\hline %
Text, Spatial & Cyber bullying detection & Word2Vec~\citep{NIPS2013_5021},AlexNet~\citep{krizhevsky2012imagenet} & \citep{Zhong2016b} \\ 
Raw ECG, ECG Spectrum & Biometrics & 1D CNN, 2D CNN & \citep{DaSilvaLuz2018} \\
\end{tabular}
\end{center}
\end{table}
}

%
%
%
\subsection{Security properties of DL algorithms}
\label{sec:sec-properties-dl-algorithms}
\paragraph{Privacy of DL Algorithms}
Although DL algorithms are successful in many domains, in some areas such as healthcare their uptake has been limited due to privacy and confidentiality concerns of the data owners. Shokri and Shmatikov proposed a system that enables multiple parties to jointly learn and use a DL model in a privacy-preserving way~\citep{ShokriRezaandShmatikov2015}. The core of their idea is to devise a novel training procedure: distributed selective stochastic gradient descent (DSSGD). In this procedure, each participant downloads the global model parameters to their local system. On their local system, they compute the gradients using stochastic gradient descent and submit a selection of the gradients to the server. Either the largest $k$ gradients select the submitted gradients, or randomly subsample $k$ gradients that are above a certain threshold $t$, where $k$ and $t$ are hyper-parameters. Their approach is not geared towards a particular DL architecture, but they validate their approach on two datasets MNIST and SVHN, with two models, an MLP and CNN. They demonstrate that the models learned by their approach guarantee differential privacy for the data owners without sacrificing predictive performance. Instead of calculating the privacy-loss per parameter, Abadi et al. calculate the privacy-loss per model to select the parameter updates and ensure differential privacy for stochastic gradient descent~\citep{Abadi2016}. The calculation of privacy-loss per model is beneficial in scenarios where a whole model will be copied to a target device, for example in the context of mobile phones. The main components of their approach are a differentially private version of the SGD algorithm, the moments accountant, and hyperparameter tuning. Abadi et al. experimentally show that differential privacy only has a modest impact on the accuracy of their baseline. %
Phan et al. propose a privacy-preserving version of a deep AE that enforces $\epsilon$-privacy by modifying the objective function~\citep{Phan2016a}. They achieve $\epsilon$-privacy by replacing the objective with a functional mechanism that inserts noise~\citep{zhang2012functional}.
Additionally, Phan et al. conduct a sensitivity analysis. Hitaj at al. develop an attack on distributed, decentralized DL and show that data of honest participants is leaked~\citep{Hitaj2017}. Their attack uses generative adversarial networks (GANs)~\citep{goodfellow2014generative} and assumes an insider at the victim's system. The insider uses a GAN to learn to create similar objects for a particular of the victim's classes and injects these images into the learning process under a different class. As a result, the victim has to reveal more gradients about the original class, thereby leaking information about the objects.   
Phong et al. demonstrate that the scheme proposed by Shokri and Shmaktikov~\citep{ShokriRezaandShmatikov2015} can leak data to an honest-but-curious server~\citep{Phong2018}. Also, they propose to address this by combining asynchronous DL with additively applied homomorphic encryption. In their evaluation setting, the computational overhead for adding homomorphic encryption to asynchronous DL is around 3.3 times the computational power of not using encryption. However, no information is leaked to the server. 

\paragraph{Adversarial Attacks on DL Algorithms}
In their work, Szegedy et al. describe that is easy to construct images that appear to be visually similar to a human observer, but will cause a DL model to misclassify~\citep{szegedy2013intriguing}. This approach requires knowledge about the parameters and architecture of the model. As an extension to that, Papernot et al. experimentally show that it is also possible to create adversarial attacks with only knowing the labels given an input~\citep{Papernot2016b}. Their black-box attack has an adversarial success rate of up to 96.19\%. Further, Papernot et al. introduce a class of algorithms that craft adversarial images~\citep{Papernot2016}, i.e., they propose an algorithm that poses a threat to the integrity of the classification. Their algorithm used the forward derivative and the saliency map to craft adversarial images and evaluated on feedforward deep neural networks and at test time they achieve an adversarial success rate of 97\%. In addition to that, Papernot et al. propose a defensive mechanism against adversarial attacks = defensive distillation~\citep{Papernot2016a}. Defensive distillation is a more robust variant of stochastic gradient descent, where additional knowledge about training points is extracted and fed back into the training algorithm. They experimentally validate that defensive distillation can reduce the effectiveness of adversarial attacks to 0.5\%. Shen et al. show that poisoning attacks are possible in a collaborative DL setting with a success rate of 99\%~\citep{Shen2016}. As a defense, Shen et al. propose a collaborative system -- AUROR -- that detects malicious users and corrects inputs. Malicious updates are detected by gradients that follow an abnormal probability distribution. AUROR reduces the poisoning attacks success rate to 3\%. Meng et al. propose MagNet, another defense system against gray-box adversarial attacks that is independent of the target classifier or the adversarial example generation process~\citep{Meng2017}. MagNet consists of two components, a detector network, and a reformer network. The detector networks is an AE that attempts to distinguish between real images and fake images, and the reformer network, which is also an AE, attempts to reconstruct the test input. Both, the detector and the reformer network are trained with Gaussian noise. Examples with larger reconstruction error of the detector network are rejected, and if they are not rejected, they are passed through the reformer to be moved closer to the original manifold to disturb their adversarial capability. During the evaluation, Magnet shows a minor reduction in classification accuracy due to the loss that is caused by the reformer network, but it is very effective to prevent adversarial attacks.

{
\small
\begin{table}
\caption{DL in InfoSec on spatial data} g
\label{tbl:sec-properties}
\begin{center}
\begin{tabular}{lccc}
\rule{0pt}{12pt} Security Property & Task & Architecture details & Approach \\[2pt]
\hline %
Privacy & Protect users privacy & custom MLP, CNN & \citep{ShokriRezaandShmatikov2015} \\ 
Privacy & Protect users privacy & custom MLP, CNN & \citep{Abadi2016} \\ 
Privacy & Protect users privacy & SDAE & \citep{Phan2016a} \\ 
Privacy & Attack privacy preserving systems & GAN\citep{goodfellow2014generative} & \citep{Hitaj2017} \\
Privacy & Protect users privacy & \citep{Phong2018} & \citep{Phong2018} \\
Integrity & Craft adversarial examples & custom MLP & \citep{Papernot2016} \\ 
Integrity & Defend against adversarial examples & custom MLP & \citep{Papernot2016a} \\ 
Integrity & Black box adversarial attacks & custom MLP & \citep{Papernot2016b} \\ 
Integrity & Defend against poisoning attacks & custom MLP & \citep{Shen2016} \\
Integrity & Defend adversarial examples & MagNet, SDAE & \citep{Meng2017} \\
\end{tabular}
\end{center}
\end{table}
}

%
%
%
%
\section{Discussion}
\label{sec:discussion}
%
%

In this survey, we have reviewed 77 papers that are about DL in the domain of InfoSec. While machine learning has played a vital role in many aspects of InfoSec for a long time, the application of DL in InfoSec is a more recent development -- 64 of the 77 papers were published after 2016.

A broad variety of different model architectures and methods have been applied in the reviewed papers. Generally, CNNs are the most common architecture for spatial data, and RNNs are the most common architecture for sequential-, structured and text data. 
The observation that multiple deep architectures can successfully be applied to the same task leads to the conjecture that the exact composition of the architecture is not the only crucial aspect for achieving good results. Often, good results are achieved by particular pre-processing or data augmentation methods, or by using more data from other sources. 
Except for these broad categorizations, it is difficult to draw a definite conclusion which architecture and solutions excel at specific tasks. In "traditional" machine learning, the main problem to solve was how to design features that model the data well. In DL, the problem shifted to choosing a suitable architecture and finding the right hyper-parameters for solving a task. 
In the DL community, there are a few empirically motivated guidelines for choosing hyper parameters~\citep{bottou2012stochastic,bengio2012practical,hinton2012practical,montavon2012neural,bergstra2012random}, as well automated attempts for hyper parameter and architecture search architecture~\citep{DBLP:journals/corr/ZophL16,bergstra2013making,bergstra2011algorithms,snoek2015scalable}. 
Within InfoSec, finding the right architecture and hyperparameters has mainly been a manual effort, or successful architectures from other domains have been adapted. 

Most of the tasks in this survey were about image or audio data. This is not surprising, because DL methods on this two data types have been hugely successful in other domains. More than one-third of the surveyed papers were about biometrics, where DL methods were used to identify individuals from sensor readings. Another prominent task was breaking CAPTCHAs, which usually involves analyzing some form of distorted images or audio clip.
Other tasks such as malware analysis or intrusion detection, that involved complex data or data that has not been studied widely outside of InfoSec have much fewer applications.

To solve InfoSec problems, 46 of them were framed as classification tasks. Using DL for classification in a supervised setting has been well researched, and many results have shown that DL methods are capable of finding solutions that generalize well to classification problems~\citep{zhang2016understanding}. The second large use case of DL methods in InfoSec is to replace manual feature engineering by automatically learned features, i.e., to learn the representations from the data. 

Only a few papers explain the results that they achieved. The other surveyed papers focus on the performance rather than on an explanation of the results. Further, lessons learned are mainly about architecture decisions. While performance is indeed important, these observations do not contribute to a better understanding of the problem or the solution. 

\subsection{Challenges}
Applying DL methods to InfoSec problems presents specific challenges. Some of these challenges exist because of additional requirements in the InfoSec domain. Others arise because of peculiarities of the data types that need to be analyzed. There are also general challenges concerning DL methods which are also of interest in InfoSec. 

One such challenge for InfoSec problems is that errors are generally costly. Consider an intrusion detection system. A false negative that is an undetected intrusion can be costly because the attacker can carry out their attack, which usually has some cost associated to it. False positives, on the other hand, are costly because a security officer needs to investigate this incident. Too many false positives may lead to many hours of effort at best, and to security officers ignoring incidents in the worst case. 

Another challenge that needs to be addressed for InfoSec problems is that in many cases the predictions of a machine learning model need to be transparent and understandable to humans, as experts need to interact with the DL models and the predictions. For example, in information forensics, the result of a model must be presentable in a court and be understandable by people there, and under some legislation, people have the right for an explanation if an automated decision has been made about them~\citep{EU2016}. Alternatively, in an intrusion detection setting, a security officer needs to understand why a specific intrusion has happened, not only that an intrusion has happened. 

Further, within the context of InfoSec, the data that is analyzed is often highly structured and carries a lot of implicit and explicit semantic meaning. For example, data that needs to be analyzed in a security context are log files, which are often structured to columns. There is no need to use DL to figure out that a character string in the date column represents a date. Outside of InfoSec, the data types where DL algorithms excel are mainly unstructured data such as the pixels of an image, text, genetic sequences or the frequency of a sound file. 

One other major challenge is that changing nature of the adversaries in information securities. Attacks and defenses evolve continually. This changing nature has to be reflected in the DL models as well. A DL model that detects malware today may potentially not be useful in the future, as the malware may change its behavior. This changing nature is very different for many other domains where machine learning is applied, where the task to be modeled remains the same. 

Then, in many cases, problems in InfoSec are in highly artificial, created contexts where lots of domain knowledge is available. For example, malware is analyzed in the context of an operating system, and the inner workings of an operating system are often available. Currently, such information is rarely used because it is challenging to combine DL models with domain knowledge. 

Finally, the DL-model that is learned is only as good that the data that has been used to build the model. Since mistakes in InfoSec are costly, ensuring high-quality data as well as high-quality labels to build the models poses another significant challenge. Labeling the data is often not trivial and labor intensive, and ensuring that a model trained on some data is qualified for a task is challenging as well. 
%
%
\subsection{Research Directions}
DL methods offer solutions to many problems that are difficult to solve by other means. However, as outlined in the previous paragraphs many challenges remain, and they are currently not very well addressed. We do believe that high performance alone is not enough and that addressing these challenges is a vital pre-condition for DL methods to be applied in a practical InfoSec setting. In this section, we outline potential research directions and point to developments in the DL community that could lead to potential solutions. 

\paragraph{Adding domain knowledge}
Currently, combining domain knowledge with DL-models poses a challenge. Domain knowledge is usually incorporated in DL methods by data pre-processing or augmenting methods, or design choices in the model architecture. DL models are built with only very general assumptions about that data, thus adding domain knowledge is likely a hard problem.

Three potential ways to add domain knowledge to DL models are the regularization of the learning process, customization of the objective functions or changes in the learning procedure. 

\paragraph{Model Adaptability}
One capability that DL models are currently lacking is model adaptability. That is, the capability to tune a model based on the judgment of an administrator. DL models can, of course, be retrained or updated with additional data, but this does not reflect the changing nature of certain threats in InfoSec. 

One-shot~\citep{vinyals2016matching,koch2015siamese,triantafillou2017few} or zero-shot~\citep{socher2013zero,johnson2016google} learning may offer a potential solution to adaptable models. Instead of learning directly to classify, these approaches learn a metric space using DL. The models learn features that distinguish certain objects in that metric space. This metric space allows new instances of entirely different objects to be described in such a space without having to re-train the model. 

Another way to achieve model adaptability is to learn DL models that represent specific, static features. Instead of modeling malware behavior using different feature types and training one model, one could train a model for each static feature type where the meaning clearly understood. An example of such a feature is modeling the number of file accesses over time using an RNN. The meaning of such a prediction is well understood - it predicts whether the amount of file accesses is within a reasonable range or unusually high. A combination of different static models could be used to build more complicated, but easy to adapt detection engines.

\paragraph{Model Interpretability}
Currently, DL methods in InfoSec focus mainly on achieving high performance for the given task. However, in many cases, the predictions of a DL model need to be understood by a human operator. Model interpretability can be understood in two ways. Either, the model and its inner workings are comprehensible, or its predictions are understandable~\citep{lipton2016mythos}. For InfoSec, we believe that the latter is more important.

Recently, work on explaining predictions of DL models has begun. Lei et al. and Riberio et al. worked on explaining predictions of classifiers~\citep{lei2016rationalizing,ribeiro2016should} by extending the model with parts that would highlight the inputs that were responsible for causing the result. Ritter et al. were inspired by methods from cognitive psychology to interpret model results and to identify biases in predictions~\citep{Ritter2017}. Montavon et al. summarize currently available tools for interpreting model predictions~\citep{montavon2017methods}. Karpathy et al. visualized the activations of RNNs to shed light on their inner workings~\citep{karpathy2015visualizing}.

Deep metric learning may offer another path to understandable representations of the data. Deep metric learning maps an input feature space to a representation feature space in such a way that a distance metric such as the Euclidean distance obtains a meaning. Deep metric learning has been successfully applied to signature~\citep{bromley1994signature} and face verification~\citep{wang2014learning,schroff2015facenet} problems as well as cross-platform code malware detection~\citep{Xu2017a}. Potentially, the learned metric space can be interpreted, though very little research has been directed towards that goal. 

Variational inference may offer yet another way to learn understandable representations. Yan et al. trained variational AE in a conditional way~\citep{yan2016attribute2image}. The resulting latent space encoded various attributes of the source images, such as objects, rotation, and shading. Bengio et al. propose a similar idea~\citep{thomas2017independently}.

Finally, reverse engineering and analyzing of the predicted results provide valuable insights on how to address problems in a particular domain. For example, DeepMind's AlphaGo, which used deep reinforcement learning to learn the game Go~\citep{silver2016mastering}. AlphaGo devised a set of previously unknown strategies to play the game, which also increased the capability of humans to play the game. Similarly, analyzing learned features may provide useful feedback on how to construct manual features or how to analyze data for particular problems.

\paragraph{Feedback Loop}
Another important aspect which is currently under-researched is how to combine the work of human experts with DL models. In particular, two questions should be addressed: How to change models after an incorrect prediction has been made, and how tune models towards certain thresholds that a human operator sets. In both cases, changes to the model need to be done in such a way that does not disturb the quality of results of other predictions. For both questions, methods of active learning~\citep{settles2012active} and changes in stochastic gradient descent may provide potential answers. 

\paragraph{Data Quality}
DL models solve tasks by deriving useful representations from data. Consequently, the quality of the model depends largely on the quality of the data. Only if the data is representative of the domain and the problems to be solved, proper models can be learned. hence, the data should be free representative of the problem domain and free from errors and biases. Bolukbasi et al. demonstrated that DL models trained on a Wikipedia text corpus would inherit the biases that are within the data~\citep{bolukbasi2016man}. In InfoSec, this may pose a significant problem. The DL community is lacking tools to decide on the quality of the underlying data, and when DL methods are used in the context of InfoSec, one of the best practices should be to use such data quality assurance tools. Research in ensuring bias-free data may also gain further traction since legislation in some parts of the world requires automated algorithms to be discrimination free, e.g. the European Union~\citep{EU2016}.

Much data in InfoSec is structured data that carries a meaning. For example, the source IP field of an IP packet intrinsically carries much meaning. However, DL methods generally work best for loosely correlated and unstructured data such as images or audio files. Consequently, two potential research directions are methods for decomposing structured data or research on novel models that are well-suited for structured data. An example for such a model is structure2vec, which uses DL techniques to learn a vector representation for structured data~\citep{song2018structure2vec}. 

High-quality labels are an essential ingredient for high-quality datasets, and in consequence for good models. Obtaining such labels and maintaining them is often labor-intensive and error-prone. Four areas of research could be pursued to mitigate the challenge of obtaining labels in InfoSec: unsupervised learning, active learning, transfer learning and metric learning. Unsupervised learning such as AE, restricted Boltzmann machines are already widely used, both within the domain of InfoSec and in other domains. Transfer learning allows for training a model on a large corpus of unrelated data, and then fine tune it to a smaller set of labeled data for the task at hand. Active learning methods allow to efficiently utilize available labeled data by carefully selecting which data is used per training batch. Moreover, finally, deep metric learning, in particular, triplet networks, may offer a solution to hard obtainable labels because they allow being trained by solving a ranking problem instead of a hard classification. Ranking data is much easier to obtain than accurate labeled data. 

\paragraph{Offensive DL}
Except for breaking CAPTCHAs, DL methods for offensive purposes are currently also rarely researched. DL-methods show significant potential to be useful in side-channel attacks. Another possibility could be to use deep reinforcement learning to train an agent that automatically attacks a system~\citep{mnih2015human}. Finally, one could use DL could be to design chatbots for phishing attacks.  

%
%
%
%
\section{Related Work}
\label{sec:related-work}
To the best of our knowledge, this is the first systematic review for DL in InfoSec from a data-centric perspective. On a broader perspective, machine learning has attracted many researchers in many sub-domains of InfoSec. In their workshop manifest, Joseph et al. provide a broad perspective on machine learning in security as well as general directions for future research~(\citep{joseph_et_al:DM:2013:4356}, ~\citep{dua2016data}). The use of machine learning was investigated in the other InfoSec sub-domains such as: intrusion detection~(\citep{sommer2010outside}, \citep{DBLP:journals/compsec/Beghdad08}, \citep{buczak2016survey}); anomaly detection~(\citep{chandola2009anomaly}, \citep{bhuyan2014network}); malware detection and classification (\citep{journals/istr/ShabtaiMEG09}, \citep{gandotra2014malware}); and information forensics~\citep{DBLP:conf/ccs/AriuGR11}.

%
%
%
%
\section{Conclusion}
\label{sec:conclusion}
This paper presents a systematic literature review on the application of DL methods in InfoSec research. We have reviewed 77 papers and presented them from a data-centric perspective, i.e., which tasks were performed on what data type. We have categorize these papers according to five different data types, sequential data, spatial data, structured data, text data as well as combinations thereof. Additionally, we have reviewed papers that have investigated the security properties of DL algorithms, such as privacy and integrity of the learning methods. 

For some well-defined issues such as biometric matching or attacking CAPTCHAs, DL methods that are successful in other domains can readily be applied and achieve state-of-the-art results. In particular, DL methods excel at machine learning tasks that are well-defined and where sufficient labeled data is available. However, many machine learning tasks in the domain of InfoSec often face a variety of unsolved challenges. Tasks are frequently hard to define, labeled data is difficult to obtain, and the data is often highly structured. Another challenge is the volatility of the machine learning tasks, e.g., attackers continuously change their behavior which is difficult to model. Besides that, DL models in InfoSec should fulfill special requirements. Domain knowledge needs to be combined with automated models, and the predictions of a model should be humanly understandable so that a security officer can judge an automated analysis and investigate on it accordingly. These requirements open ample opportunities for research, such as adapting and tuning models, combining domain knowledge with models, transparency of the models or learning to understand the predictions of DL methods.

To conclude, we want to re-iterate one of the most significant merits of DL methods, namely: DL methods derive useful representations from data, which leads to two very desirable consequences. First, it saves the manual effort that was previously required to manually craft features, especially in domains where it is difficult to hand-craft features because the domain is hard to understand. Secondly, it allows models and methods that are successful on a specific data type of one domain be also applicable to other problems of other domains on a similar data type. These two merits potentially lead to synergies between domains that have previously been disconnected, for example, health care and InfoSec. Advances in one domain, e.g., on the transparency of the models can readily be deployed in other domains.  

\section*{Acknowledgment} 
The work presented in this paper is part of a project which has received funding from the European Union’s Horizon 2020 research and innovation programme under grant agreement No 780495. 


\bibliographystyle{plainnat} 
\bibliography{references}

\appendix
\section{Venues}
\label{app:venues}
As indicated in Section \ref{sec:survey-methodology}, here  we list all the venues that we included in our review. The venues are listed in alphabetical order. 

\paragraph{Security venues}
We included literature of the following security conference proceedings and journals: ACNS, ACSAC, ARES, ASIACCS, ASIACRYPT, CCS, Computers and Security, CRYPTO, ESORICS, EUROCRYPT, Fast Software Encryption, FC, IACR, ICB, ICC, IEEE Transactions on Dependable and Secure Computing, IWSEC, Journal of Cryptology, Privacy Enhancing Technologies Symposium, RAID, S\&P, SACMAT, SIGCOMM, SOUPS, Theory of Cryptography, Transactions on Information Forensics and Security, TrustCom, WISEC, WPES.

\subsection{Machine learning venues}
We included literature of following machine learning conference proceedings and journals: ACCV, ACM SIGKDD, ACM Transactions on Intelligent Systems and Technology, ACM Transactions on Knowledge Discovery from Data, Advances in Data Analysis and Classification
AISTATS, BioData Mining, BMVC, Computer Vision and Image Understanding, CVPR, Data Mining and Knowledge Discoveries, ECCV, ECML PKDD, ICCV, ICDM, ICIP, IEEE Computer Society Conference on Computer Vision and Pattern Recognition, IEEE International Conference on Big Data, IEEE Transactions on Image Processing, IEEE Transactions on Knowledge and Data Engineering, IEEE Transactions on Pattern Analysis and Machine Intelligence, Image and Vision Computing, International Conference on Pattern Recognition, International Journal of Computer Vision, Journal of Visual Communication and Image Representation, Knowledge and Information Systems, Machine Vision and Applications
Medical Image Analysis, PAKDD, Pattern Recognition, Pattern Recognition Letters, RecSys, SDM, shops, Social Network Analysis and Mining, Statistical Analysis and Data Mining, WACV, WSDM

\end{document}